\documentclass{jfm}
\usepackage{graphicx}
\usepackage{amsmath, amssymb}
\usepackage[inactive,tightpage]{preview}
\usepackage{microtype}
\usepackage[normalem]{ulem}
\usepackage{mathtools,bm}
\usepackage{esvect}
\usepackage{enumitem}
\usepackage{import}
\usepackage{mdframed}

\pdfmapfile{+rsfso.map}
\DeclareSymbolFont{rsfso}{U}{rsfso}{m}{n}
\DeclareSymbolFontAlphabet{\mathscr}{rsfso}

\usepackage{pdflscape}
\usepackage{afterpage}
\usepackage{flafter}
\usepackage{rotating}
\usepackage{fancyhdr}
\fancypagestyle{lscape}{% 
\fancyhf{} % clear all header and footer fields 
\fancyfoot[LE]{}
\fancyfoot[LO] {}
 
}

\usepackage{bm}
\usepackage{microtype}

\usepackage{hyperref}
\usepackage[usenames, dvipsnames]{xcolor}
\hypersetup{colorlinks=true,linkcolor=MidnightBlue,citecolor=MidnightBlue}
\usepackage{natbib}
\usepackage{array}
\usepackage{booktabs}
\usepackage{multirow}
\usepackage{cleveref}
\usepackage{diagbox}

\usepackage[labelfont=sc]{caption}
\usepackage[position=top]{subcaption}
\usepackage{tabularx}
\usepackage{siunitx}
\newcolumntype{Y}{>{\centering\arraybackslash}X}

\usepackage{tikz}
\usepackage{pgfplots}
\usetikzlibrary{backgrounds, calc, patterns.meta, shadows, arrows,positioning,shapes,fit}

\usepackage{pdflscape}
\usepackage{afterpage}
\usepackage{rotating}
\usepackage{fancyhdr}
\fancypagestyle{lscape}{% 
\fancyhf{} % clear all header and footer fields 
\fancyfoot[LE]{}
\fancyfoot[LO] {}
 
}

\usepackage[outline]{contour} % For outlining fonts
\contourlength{0.15em}
\usetikzlibrary{external}

\usepackage{grffile}
\pgfplotsset{compat=newest}
\usetikzlibrary{plotmarks}
\usetikzlibrary{arrows.meta}
\usepgfplotslibrary{patchplots}
\usepgfplotslibrary{fillbetween}

\usepackage[outline]{contour} % For outlining fonts
\contourlength{0.15em}

\usepackage{lscape}

\renewcommand{\P}{P}
\renewcommand{\r}{r}

\renewcommand{\t}{\ell}
\newcommand{\K}{K}
\newcommand{\timeA}{\mathrm{A}}

\newcommand*{\ep}{\epsilon}

\renewcommand*{\i}{\mathrm{i}}
\newcommand*{\im}{\mathrm{i}}
\newcommand*{\e}{\mathrm{e}}

\renewcommand*{\Re}{\operatorname{Re}}
\renewcommand*{\Im}{\operatorname{Im}}

\newcommand*{\de}{\operatorname{d\!}{}} % Do this
\newcommand{\dd}[2]{\frac{\de#1}{\de#2}}

\newcommand{\pd}[2]{\frac{\partial#1}{\partial#2}}

\newcommand{\myblue}[1]{{\color{black}#1}}

\usepackage{printlen}

\def\Xint#1{\mathchoice
   {\XXint\displaystyle\textstyle{#1}}%
   {\XXint\textstyle\scriptstyle{#1}}%
   {\XXint\scriptstyle\scriptscriptstyle{#1}}%
   {\XXint\scriptscriptstyle\scriptscriptstyle{#1}}%
   \!\int}
\def\XXint#1#2#3{{\setbox0=\hbox{$#1{#2#3}{\int}$}
     \vcenter{\hbox{$#2#3$}}\kern-.5\wd0}}
\def\YYint#1#2#3{{\setbox0=\hbox{$#1{#2#3}{\int}$}
     \vcenter{\hbox{\scalebox{1}[-1]{$#2#3$}}}\kern-.5\wd0}}
\def\(\alpha\)dashint{\Xint=}
\def\dashint{\Xint-}

\shorttitle{Exponential asymptotics for Saffman-Taylor in a wedge}
\shortauthor{Andersen, Lustri, McCue \& Trinh}

\title{On the selection of Saffman-Taylor viscous fingers for divergent flow in a wedge}

\author{Cecilie Andersen$^1$, Christopher J. Lustri$^2$, \\ Scott W. McCue$^3$  
 \and Philippe H. Trinh$^{1,4}$}

\affiliation{
    $^1$Department of Mathematical Sciences, University of Bath, Bath BA2 7AY, UK\\
    
    $^2$School of Mathematics and Statistics, The University of Sydney, Sydney NSW 2006, Australia\\
    
    $^3$School of Mathematical Sciences, Queensland University of Technology, Brisbane QLD 4001, Australia \\
    $^4$Theoretical Sciences Visiting Program (TSVP), Okinawa Institute of Science and Technology Graduate University, Onna, 904-0495, Japan

}

\pubyear{XXXX}
\volume{YYY}
\pagerange{xxx--xxx}
\date{\today~[Draft]}

\makeatletter
\newcommand\Label[1]{&\refstepcounter{equation}(\theequation)\ltx@label{#1}&}
\makeatother
\begin{document}

\maketitle

\begin{abstract}
We study self-similar viscous fingering for the case of divergent flow within a wedge-shaped Hele-Shaw cell. Previous authors have conjectured the existence of a countably-infinite number of selected solutions, each distinguished by a different value of the relative finger angle. Interestingly, the associated solution branches have been posited to merge and disappear in pairs as the surface tension decreases. 
We demonstrate how exponential asymptotics is used to derive the selection mechanism. In addition, asymptotic predictions of the finger-to-wedge angle are given for different sized wedges and surface-tension values. 
The merging of solution branches is explained; this feature is qualitatively different
to the case of classic Saffman-Taylor viscous fingering in a parallel channel configuration. The phenomena of branch merging in our self-similar problem relates to tip splitting instabilities in time-dependent flows in a circular geometry, where the viscous fingers destabilise and divide in two.

%The merging of solution branches is explained; this feature is unique in comparison to the case of classic Saffman-Taylor viscous fingering in a parallel channel configuration. For the case of time-dependent flows in a circular geometry, the phenomena of branch merging relates to tip-splitting instabilities, where the viscous fingers destabilise and divide in two.
%
% Previous results have demonstrated that
% %
% The resulting bifurcation diagram of relative finger angle against surface tension shows a countable family of selected solutions which merge and disappear in pairs as the surface tension decreases. 

\end{abstract}

%\begin{preview}
%\import{FINALFIG/}{graphical_abstract_2.pdf_tex}
%\end{preview}

\section{Introduction}

\begin{figure}
    \centering
    %\begin{preview}
    %\input{TIKZ/BifurcationDiagram_channelandvsBA}
    %\end{preview}
    \includegraphics{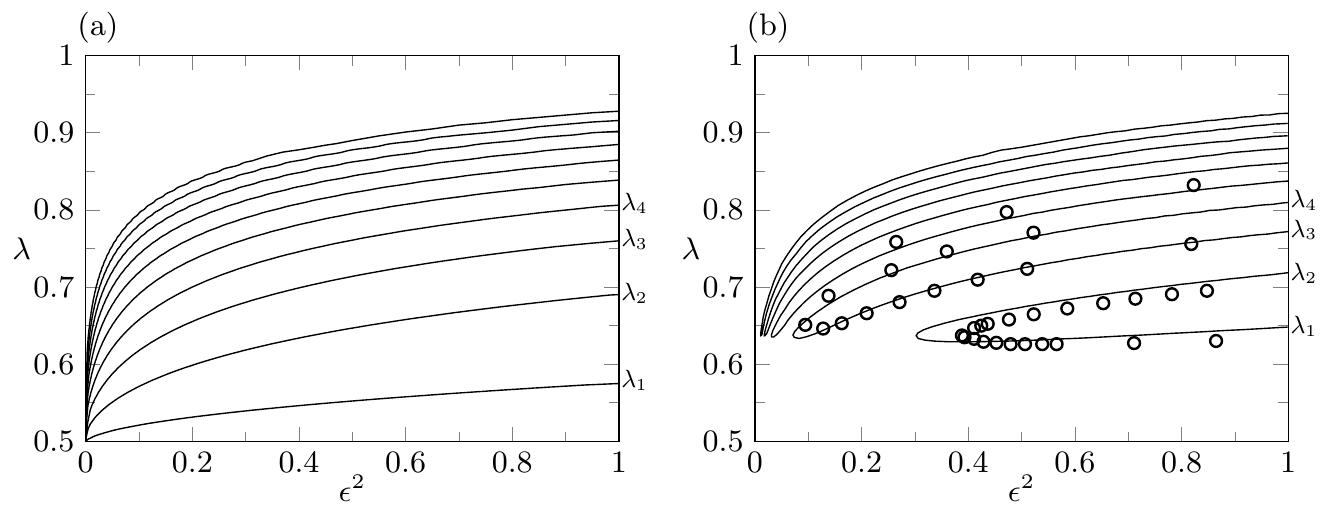}
    \caption{(a) The bifurcation diagram for the classic Saffman-Taylor problem in a channel showing the first 10 branches of selected $\lambda$ values against the surface tension parameter $\epsilon^2$. \myblue{Labels are shown for the first four branches: $\lambda_1$, $\lambda_2$, $\lambda_3$, $\lambda_4.$ This is equivalent to the results of \cite{chapman1999}.} (b) The solid lines show the bifurcation diagram which we calculate using exponential asymptotics for wedge angle $\theta_0 = 20^\circ$. \myblue{This shows the permitted $\lambda(\epsilon)$ values that are selected by the selection mechanism. Details of the derivation of this selection mechanism follow in rest of the paper.} The circles show the numerically calculated values extracted from \cite{amar1991}. Notice that $\lambda_1$ and $\lambda_2$ merge at $\epsilon \approx 0.3$ and $\lambda_3$ and $\lambda_4$ merge at $\epsilon \approx 0.07$.}
    \label{fig:BAvsourBifurcationDiagram}
\end{figure}

\noindent In their now classic work on viscous fingering, \citet{saffmantaylor1958} consider the situation of a single finger of air---or an otherwise less viscous substance---steadily penetrating a Hele-Shaw cell filled with a viscous fluid. A key quantity of interest is the proportion of the channel occupied by the width of the finger, denoted $\lambda \in (0,1)$. Experimentally, Saffman and Taylor observed that $\lambda \approx 1/2$. However, their asymptotic analysis, valid at small values of the non-dimensional surface tension parameter $\epsilon \to 0$, did not seem to restrict the value of $\lambda$. Today, it is known that for a fixed $\epsilon$ there exists a countably infinite number of possible values of $\lambda = \lambda_i$ with the property that
\begin{equation} \label{eq:selectionST}
1/2 <\lambda_1(\epsilon)<\lambda_2(\epsilon)< \cdots <1. 
\end{equation}
In the limit $\epsilon \to 0$, every element in the family converges to 1/2 (see \emph{e.g.} \citealt{VBbook}). The resultant plot of the eigenvalue, $\lambda$, versus the surface tension parameter, $\epsilon$, is shown in \cref{fig:BAvsourBifurcationDiagram}(a). The resolution of the Saffman-Taylor problem, including the asymptotic derivation of \eqref{eq:selectionST} is obtained using exponential asymptotics; today it is a prototypical example of such beyond-all-orders asymptotics [see \emph{e.g.} works by \citealt{Mclean1981, Hong1986, Combescot1986, Combescot1988, Tanveer1987, Tanveer2000, chapman1999, SaffmanReview}].

\begin{figure}
    \centering
    %\begin{preview}
    %\input{TIKZ/WedgeinCircularGeometryTikz}
    %\end{preview}
    \includegraphics{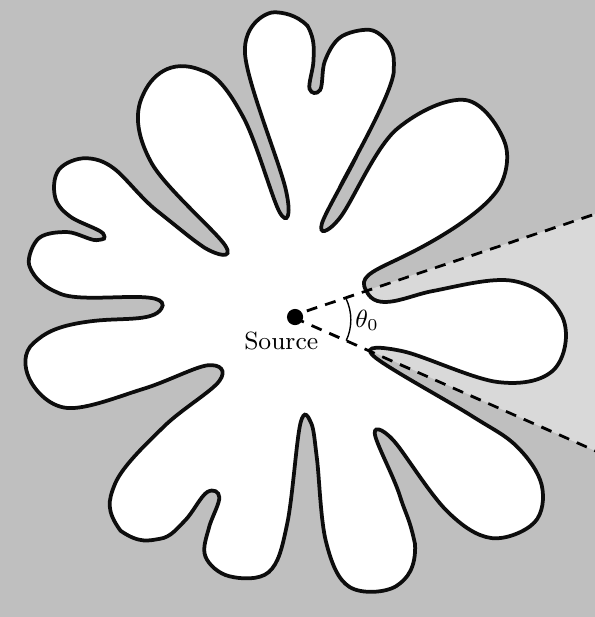}
    \caption{Sketch of the free surface for viscous fingering in the full circular geometry, where a central source injects fluids outwards in all directions. Assuming axi-symmetry, a single finger from this free surface can be considered as arising due to injecting fluid in the corner of Hele-Shaw cell limited by side walls consisting of a wedge of angle $\theta_0$.}
    \label{fig:wedgefromcircular}
\end{figure}

In this paper, we are interested in deriving a similar selection law to \eqref{eq:selectionST} but for an analogue problem where fluid is injected at the corner of a Hele-Shaw cell limited by side walls consisting of a wedge of specified angle. Although this wedge-scenario is interesting in its own right, it gains further importance as a partial model for the fingering seen where fluid is injected into a Hele-Shaw cell from a central source. As the injected fluid moves outwards, the interface destabilises and fingering occurs within the manner illustrated in \cref{fig:wedgefromcircular}. Thus, the limited wedge configuration considered in this work serves as a model for each sector of the full source problem. 

Consider the wedge problem characterised by a wedge angle, $\theta_0$, measured at the corner. The key parameter, $\lambda$, now describes the angular proportion, $\lambda\theta_0$, of the cell occupied by the finger. In comparison with the previous \cref{fig:BAvsourBifurcationDiagram}(a) for classic Saffman-Taylor viscous fingering, in \cref{fig:BAvsourBifurcationDiagram}(b) we plot the bifurcation diagram for the case of a wedge of angle $\theta_0 = 20^\circ$. The solid lines correspond to the new asymptotic predictions developed in this work, while the circles correspond to the previous numerical results of \cite{amar1991}. The figure shows the existence of distinct solution families, $\lambda = \lambda_i(\ep, \theta_0)$. 

In addition to the expected solution families, $\lambda_i$, there is now an additional \myblue{phenomenon} for the wedge-limited configurations. As seen in \cref{fig:BAvsourBifurcationDiagram}(b), there exist certain critical values of the surface tension parameter, $\ep$, where each solution family, $\lambda_i$, \myblue{reaches} a turning point connected to the adjacent family, i.e. $\lambda_i = \lambda_{i+1}$. In this work, we will explain how this merging of eigenvalues can be understood from the perspective of the exponential asymptotics. As noted by \citet{amar1991}, the merging of eigenvalue pairs causes a loss of existence of the solutions for sufficiently small $\epsilon$. Physically, in connection with the full geometry in \cref{fig:wedgefromcircular}, this results in the finger splitting into two through a tip splitting instability. One should then consider two wedges of half the angle to continue to follow the finger profiles. It is then interesting to consider the consequences of the exponential asymptotic analysis towards the more complicated problem of time-dependent tip splitting instabilities in the unrestricted planar domain. This will be further discussed in \cref{sec:discuss}.

% We shall see that there exist certain thresholds of the surface tension where each solution family, $\lambda_i$, eventually terminates at a turning point connected to the adjacent family. These transition thresholds depend on the wedge angle: for example, for a wedge angle of $20^\circ$ \cref{fig:BAvsourBifurcationDiagram}(b) shows that $\lambda_1$ and $\lambda_2$ merge at $\epsilon \approx 0.41$ and then $\lambda_3$ and $\lambda_4$ merge at $\epsilon \approx 0.1$. In this problem the eigenvalues $\lambda_i$ depend both on the surface tension and the wedge angle. In this paper we will derive semi-analytical formulae for the selection mechanism. An example of the selected families of solutions is shown in \cref{fig:BAvsourBifurcationDiagram} for a wedge angle of $20^\circ$. In the figure, our analytical results are compared against the numerical results from \cite{amar1991} and our results agree qualitatively well. 

% Future work would aim to extend this analysis to the full circular geometry where a central source injects fluid outwards in all directions.

\subsection{Background and open challenges of the wedge problem}

\noindent The literature on Hele-Shaw flows and viscous fingering problems is extensive; here we provide a review of selected papers, primarily focused on wedge configurations or closed bubbles in channels. 

Experimental observations and initial analysis for the wedge geometry can be found in the works of \cite{paterson1981} and \cite{thome1989}. Then in the early 1990s, a number of works appeared on the beyond-all-orders aspects of the wedge configuration \myblue{(\citealt{brener1990}, \citealt{amar1991exact}, \citealt{amar1991}, \citealt{tu1991},  \citealt{combescot1992},  \citealt{LevineTu1992}).} \cite{combescot1992} identified the singularities in the complex plane that are responsible for the selection mechanism. The solvability condition was found exactly by \cite{brener1990} for the case with a $90^{\circ}$ separation angle, where the solution reduces to a closed analytic form. Both \cite{combescot1992} and \cite{tu1991} used WKB methods to make analytic progress on the problem with a general wedge angle whilst numerical results were obtained by \cite{amar1991}.

This paper is most strongly motivated by the work of \cite{tu1991} and \cite{amar1991}. \cite{tu1991} had linearised the free-surface problem with a general wedge angle to obtain a model differential equation. For this new equation, a WKBJ (Liouville-Green) approximation was used to derive a solvability condition that can predict the theoretical zero-surface tension limit for $\lambda$, as well as a condition that finds the point in the bifurcation diagram where branch merging occurs.

\cite{amar1991} produced accurate numerical results for parts of the bifurcation diagram. However, on account of the challenges in numerical computations of the eigenvalue problem, \cite{amar1991} noted that:
\begin{quotation}
\noindent \emph{Our predictions concerning levels [branches] higher than the first two require confirmation by a very careful WKB analysis, which is the most suitable treatment at extremely low surface tension. Probably, the results of analytic solutions without surface tension will make this analysis possible.}
\end{quotation}
At that point (and until the present) we do not believe any group has managed to derive the exact selection mechanism (i.e. the missing analysis referenced above).

Modern techniques of exponential asymptotics \citep{Chapman1998} allow us to study the wedge problem in the small surface tension limit without the need to linearise in the same fashion as previous authors. In this paper, we will use these techniques to address the open problem identified by \cite{amar1991} and obtain an analytic solvability condition for the selected eigenvalues.

More recently, further work has been done on closely related problems in Hele-Shaw channels. There is particular interest in Hele-Shaw channels with a central raised rail, which can change the stability of the finger \citep{FrancoGomez2016, Thompson2014}. Further, there have been many recent experimental \citep{Lawless2023, FrancoGomez2017, Gaillard2021}, numerical \citep{Thompson2021, FrancoGomez2017, Keeler2019} and analytic \citep{Booth2023, Keeler2019} developments concerning closed bubbles propagating along Hele-Shaw channels. We will return at the end of the paper to discuss the connection of our work with these newer problems.

\section{Mathematical formulation}

\begin{figure}
    \centering    \includegraphics{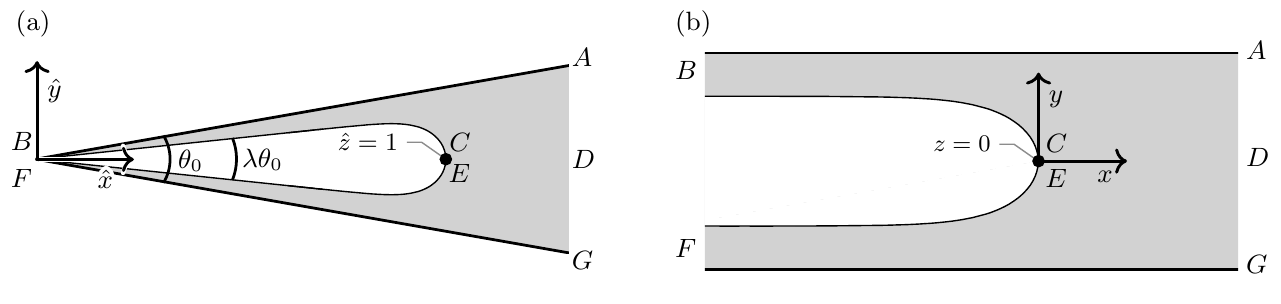}
    %\begin{preview}
    %\input{TIKZ/PhysicalProfile}
    %\end{preview}
            \caption{(a) A numerical plot of the top-down view of the self-similar physical profile in the $\hat{z}$-plane is shown for the zero surface tension case, with parameter values $\theta_0 = 20^{\circ}$ and $\lambda = 0.6$. The free-surface was computed using \eqref{eq:xyhypergeometric}.  The Hele-Shaw cell is bounded by the thick black lines and is filled with a viscous fluid shown in grey. An inviscid fluid is injected from the corner of the wedge and forms a finger with angle $\lambda\theta_0$. The corner of the wedge lies at $\hat z = 0$ ($BF$) and the tip of the finger lies at $\hat z=1$ ($CE$).  \myblue{(b) A sketch of the $z$-plane, computed via the conformal map $z=(2/\theta_0)\log\hat{z}$.  This configuration is analogous to the traditional Saffman-Taylor finger (\citealt{Mclean1981}).}}
    \label{fig:HeleShaw}
\end{figure}

\noindent \myblue{A traditional Hele-Shaw cell consists of two parallel plates separated by a small distance, $b$, and filled with viscous fluid with viscosity, $\mu$. For the case of a circular geometry, like that depicted in \cref{fig:wedgefromcircular}, an inviscid fluid is injected at a point, and this drives an outwards-expanding interface between viscous and inviscid fluids. 

We now consider the related wedge-shaped geometry, as shown in \cref{fig:HeleShaw}(a). Here the inviscid fluid is injected at the wedge corner at some prescribed flow-rate. The viscous fluid is constrained to lie between the wedge walls separated by an internal angle $\theta_0$. As can be observed experimentally \citep{thome1989}, a self-similar shape is reached eventually, where the inviscid fluid occupies an angle $\lambda \theta_0$, with $0<\lambda<1$. This set up is referred to as \emph{divergent flow} in \cite{amar1991exact}. For the case of zero surface tension, a prototypical solution is shown in \cref{fig:HeleShaw}(a), and we observe the petal-shaped interface between viscous fluid and inviscid fluid.}  

%Consider a Hele-Shaw cell with a very small thickness compared to the length so the thickness of the fluid can be considered to be negligible. The cell is in the shape of a wedge and we choose $\theta_0$ to denote the internal angle of the wedge, for consistency with notation in \cite{amar1991exact}. The wedge is filled with a viscous fluid with viscosity $\mu$. An inviscid fluid is injected from the corner of the wedge at some prescribed flow-rate and displaces the viscous fluid to form a petal/finger shape. Eventually, as can be observed experimentally \citep{thome1989}, a self-similar shape is reached where the petal occupies an angle $\lambda \theta_0$, where $0<\lambda<1$. This set up is referred to as `divergent flow' in \cite{amar1991exact}; an example profile is presented in \cref{fig:HeleShaw}.

The classic problem of Saffman-Taylor viscous fingering in a channel is typically studied in a travelling frame corresponding to a steady-state finger. In the wedge geometry, the analogue of a travelling wave frame of reference is a self-similar solution. \myblue{In Appendix~\ref{app:formulation}, we demonstrate how the original equations of potential flow for a Hele-Shaw cell, with boundary conditions on the channel walls and free-boundary, and injection condition can be reposed in the self-similar framework. The key idea relates to a transformation of the original dimensional lengths, $\bar{x}$ and $\bar{y}$, which are scaled via} 
\begin{equation}
(\hat{x}, \hat{y}) = \frac{(\bar{x}, \bar{y})}{R_0 A(t)},
\end{equation}
\myblue{where $R_0$ is a length scale (chosen to be the distance between the corner of the wedge and the tip of the finger at $t=0$) and $\mathrm{A}(t)$ is a dimensionless scaling factor that depends on dimensionless time, $t$. As shown in Appendix~\ref{app:formulation}, two possible choices of $\mathrm{A}$ result in effectively the same self-similar problem, given by \eqref{nearfinal_nondim}, in dimensionless variables with a time-independent effective surface tension parameter $\sigma$. This $\sigma$, when rescaled, then gives us our small parameter $\epsilon$, defined below in \eqref{eq:epsilontosigma}; see \citealt{amar1991} for further details}.

% The first way to derive our self-similar problem is to set $\mathrm{A}(t)^2\mathrm{A}'(t)=1$ and allow the flow rate at infinity to scale like $1/\mathrm{A}$ (i.e., $(3t+1)^{-1/3}$).  While not typically employed in practice, such a time-dependent flow rate can be realised experimentally.  The second way is to set $\mathrm{A}(t)\mathrm{A}'(t)=1$, keep the flow rate at infinity constant, but allow the surface tension to scale like $1/\mathrm{A}$ (i.e., $(2t+1)^{-1/2}$).  This option is less obvious from a physical perspective, as surface tension is a constant physical quantity; however, \cite{amar1991exact} argues that this derivation is reasonable as the time-dependence is slowly varying.  

%The physical plane, given by $z = x + \im y$, is shown in \cref{fig:HeleShaw}. We assume that the fluid is incompressible and irrotational, and therefore possesses a complex potential, denoted $f = \phi+ \i\psi$, which satisfies Laplace's equation, $\nabla^2 \phi = 0$. Within the $f$-plane, the problem has been re-scaled so that the fluid is located in the infinite strip bounded by $\psi \in [-1,1]$.

% {\color{magenta}We very briefly summarise the complex variable formulation of the problem here and refer the reader to \cite{amar1991exact} for a full explanation.}

\myblue{We thus have the governing set of potential-flow equations in \eqref{nearfinal_nondim} corresponding to a scaled velocity potential, $\hat{\phi}$, and self-similar physical-plane coordinates, written in complex form as $\hat z = \hat x + \im \hat y$ [\cref{fig:HeleShaw}(a)]. Note that the free-surface is now stationary. We introduce the complex potential, $\hat{f}=\hat{\phi}+\i\hat{\psi}$ with harmonic conjugate, the streamfunction $\hat\psi$.  Within the $\hat{f}$-plane, fluid is located in the infinite strip bounded by $\hat\psi \in [-1,1]$. 

Finally, the flow domain is mapped to a channel geometry via \begin{equation}
z= \frac{2}{\theta_0}\log \hat{z},
\end{equation}
so that the walls $BA$ and $FG$ lie on $\Im z=\pm 1$, respectively, and the tip $CE$ is fixed to the origin $z=0$. This is shown in \cref{fig:HeleShaw}(b).}

%change to appendix in revised version
\myblue{Following section 3B of \cite{amar1991}, we review the procedure for developing a set of boundary-integral equations for the potential-flow problem.} First, the velocity potential is shifted as
\begin{equation}
f^*=\phi^*+\i\psi^*=\frac{\hat{f}-H(z)}{(1-\lambda)Q_0/\lambda}.
\end{equation}
Above, $2Q_0$ is the dimensionless flux of fluid \myblue{across the wedge at infinity in the self-similar frame [cf. later \eqref{eq:Q0}]}. The function $H(z)$ is implicitly defined so that the free surface lies on $\psi^*=$ constant \myblue{[cf. eqn (3.10) of \cite{amar1991}]. For the case of the classic Saffman-Taylor viscous fingering problem in a parallel channel, $H(z) = z$, as shown by \cite{Mclean1981}.}

\begin{figure}
    \centering
    %\begin{preview}
    %\input{TIKZ/ZetaPlane}
    %\end{preview}
    \includegraphics{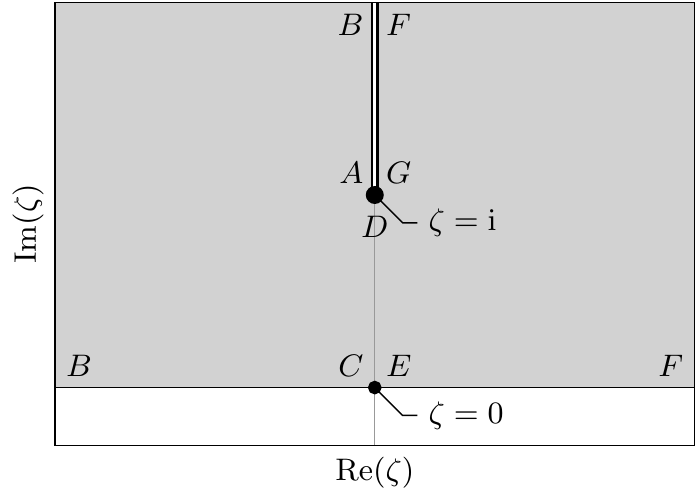}
    \caption{The fluid region (grey) is mapped to the upper-half $\zeta$-plane. The key points from \cref{fig:HeleShaw} are labelled here. Within the $\zeta$-plane, there is a branch cut from the point $D$ ($\zeta = \i$). Here the branch cut is taken vertically up the imaginary axis from $\zeta = \i$.}
    \label{fig:ZetaPlane}
\end{figure}

As in \cite{chapman1999}, it is convenient for later analysis to map the fluid region to the upper-half $\zeta$-plane via
\begin{equation}
\zeta^2 = \e^{\pi f^*} - 1.
\label{eq:ftozetamap}
\end{equation}
The mapped fluid domain for the configuration in \cref{fig:HeleShaw} is shown in \cref{fig:ZetaPlane}. Thus we see that under the map \eqref{eq:ftozetamap}, the free surface, denoted $BCEF$, lies on the real $\zeta$-axis, while the tip of the finger, denoted, $CE$, is at $\zeta = 0$. 

%obtaining a steady reference frame is more complicated. It was shown experimentally by \cite{thome1989} that after a given time the finger reaches a self-similar shape, but with a degree of surface tension that varies slowly with time. We therefore study the finger in this self-similar regime, where it is possible to fix a coordinate system which moves with the finger. In this frame, we obtain a re-scaled \emph{effective surface tension} which is time-independent, as in \cite{amar1991}. We will denote this modified surface tension parameter by $\sigma$.

In the governing equations to follow, we shall seek to solve for the unknown free surface location and fluid velocities along the interface, $\zeta = \xi \in \mathbb{R}$. It is convenient to introduce quantities $q$ and $\tau$ via
\begin{equation} \label{eq:qtaudef}
\frac{q}{1-\lambda}\e^{-\i\tau} = \dd{f^*}{z} = \pd{\phi^*}{x} + \i \pd{\phi^*}{y},
\end{equation}
which are analogues of speed and streamline angle, respectively (and reduce to the actual fluid speed and streamline angle in the limit $\theta_0\rightarrow 0$). Therefore,  we require a set of governing equations for the unknowns $(x(\xi), \, y(\xi), \, q(\xi), \, \tau(\xi))$.

%the speed, $q$,  and streamline angle, $\tau$, via 
%\begin{equation} \label{eq:qtaudef}
%q\e^{-\i\tau} = \dd{f}{z} = \pd{\phi}{x} + \i \pd{\phi}{y},
%\end{equation}
%and therefore, we require a set of governing equations for the unknowns $(x(\xi), \, y(\xi), \, q(\xi), \, \tau(\xi))$.

\myblue{With the various conformal maps now established, at this point, we may follow the same procedures as found in section 3B of \cite{amar1991}}. We find on the free surface, where $\zeta = \xi$ is real, that continuity of pressure yields Bernoulli's equation:
\begin{equation}
\begin{split}
\epsilon^2  \frac{P(\xi)}{4} \frac{\partial}{\partial \xi}\left(r(x) \left[-P(\xi) q(\xi)\frac{\partial \tau}{\partial \xi} + \frac{\t}{2}\sin\tau(\xi)\right]\right) \\= 
\frac{Q_0}{1-\lambda} - \frac{2}{\pi}(1+\xi^2)\dashint_{-\infty}^0 \frac{\K(\tilde{\xi}) \cdot \de{\tilde{\xi}}}{P(\tilde{\xi})(\xi^2- \tilde{\xi}^2)},
\label{eq:bernoulliequation}
\end{split}
\end{equation}
\myblue{which can be compared to eqn (3.10) of \cite{amar1991}.} In \eqref{eq:bernoulliequation}, we have defined a number of functions for convenience. Firstly, we have written
\begin{equation} \label{eq:PrK}
\begin{gathered}
\P(\xi) = (1+\xi^2)/\xi, \qquad
\r(x) = \e^{-\theta_0 x(\xi)/2}, \qquad
\K(\xi) = \frac{\sin [\tau(\xi)] }{q(\xi) [r(x(\xi))]^2}.
\end{gathered}
\end{equation}
Note that $r$ is a function of $x$ which, in turn, depends on $\xi$. In \eqref{eq:bernoulliequation}, we have defined $Q_0$ to be
\begin{equation} \label{eq:Q0}
Q_0 =  \frac{2(1-\lambda)}{\pi} \int_{-\infty}^0 \frac{\K(\tilde{\xi})}{P(\tilde{\xi})} \, \de{\tilde{\xi}},
\end{equation}
which, as \myblue{mentioned} above, represents a dimensionless constant describing the fluid flux.  It is also convenient to define a scaled value for the interior wedge angle $\theta_0$,
\begin{equation}
    \t = \t(\lambda) = \frac{\theta_0}{\pi} (1 - \lambda),
    \label{eq:elldefinition}
\end{equation}
where $\lambda$ is the proportional finger angle parameter. Finally, we have introduced the key non-dimensional parameter, $\epsilon$, by 
\begin{equation}
\epsilon^2  = \frac{4\pi^2 \hat{\sigma}}{(1-\lambda)^2},
\label{eq:epsilontosigma}
\end{equation}
where $\hat{\sigma}$ is the modified surface tension parameter \myblue{presented eqn \eqref{eq:benamar_paramchoice} in Appendix \ref{app:formulation}}. 
%from \cite{amar1991}.
We consider $\epsilon^2$ to be a small parameter, corresponding to the small surface-tension regime, and we will therefore study the problem in the asymptotic limit $\epsilon \rightarrow 0$.

Analyticity of $q\e^{-\im \tau}$ in the upper-half $\zeta$-plane gives, by the Hilbert transform,
\begin{equation}
\log q(\xi) = \mathcal{H}[\tau](\xi) \quad \text{where} \quad \mathcal{H}[\tau](\xi) = \frac{2}{\pi}\dashint_{-\infty}^0 \frac{\tau(\tilde{\xi}) \, \tilde{\xi}}{\xi^2 - \tilde{\xi}^2} \, \de{\tilde{\xi}},
\label{eq:hilberttransformequation}
\end{equation}
where we have defined the operator, $\mathcal{H}$. Finally, we close the system by integrating the free surface velocity relationships \eqref{eq:qtaudef}. This yields
\begin{equation}
x(\xi) +\i y(\xi) = \frac{2(1-\lambda)}{\pi}\int_\xi^0 \frac{\e^{\i\tau(\tilde{\xi})}}{q(\tilde{\xi}) P(\tilde{\xi})} \, \de{\tilde{\xi}}.
\label{eq:xyequation}
\end{equation}

Thus, the full system consists of equations (\ref{eq:bernoulliequation}), (\ref{eq:hilberttransformequation}) and (\ref{eq:xyequation}) for the unknowns $(x, y, q, \tau)$ \myblue{in addition to the boundary conditions,
\begin{subequations}
    \begin{align}
        &x(\pm \infty) = -\infty, \quad y(\pm \infty) = \mp \lambda, \\
        &q(\pm \infty) = 1, \quad \tau(-\infty) = 0, \quad \tau(\infty) = -\pi.
    \end{align}
\end{subequations}} \myblue{Above, the first set of boundary conditions correspond to imposing the geometrical constraints while the second set correspond to the velocity and streamline angle constraints. In total, the equations and boundary conditions are equivalent to those of \cite{amar1991exact}.} In the next section, we shall examine the zero surface-tension solutions of these equations.

\section{Zero surface-tension solutions}

\noindent We first discuss the zero surface-tension solutions, $(x_0(\xi), y_0(\xi), q_0(\xi), \tau_0(\xi)),$ which correspond to setting $\epsilon = 0$ in the governing equations (\ref{eq:bernoulliequation}), (\ref{eq:hilberttransformequation}) and (\ref{eq:xyequation}). We thus approximate $x \sim x_0$, $y \sim y_0$, $q \sim q_0$, and $\tau \sim \tau_0$ in the limit $\ep \to 0$. The zero surface-tension equations are given by,
\begin{subequations} \label{eq:govO1}
\begin{align}
     0 &=\frac{Q_0}{1-\lambda} - \frac{2}{\pi} (1+\xi^2)\dashint_{-\infty}^0 \frac{K_0(\tilde{\xi}) }{ P(\tilde{\xi})(\xi^2-\tilde{\xi^2})}\de{\tilde{\xi}}, \\
    \log q_0(\xi) &= \mathcal{H}[\tau_0](\xi), \label{eq:hilberttransformleadingorder} \\
    x_0(\xi) + \i y_0(\xi) &= - \frac{2(1-\lambda)}{\pi}\int_\zeta^0 \frac{\e^{\i\tau_0(\tilde{\xi})}} {q_0(\tilde{\xi})P(\tilde{\xi})}\de{\tilde{\xi}},\label{eq:xyleadingorder}
\end{align}
\end{subequations}
\myblue{where,}
\begin{equation}
K_0(\xi) = \frac{\sin[\tau_0(\xi)]}{q_0(\xi)[r(x_0(\xi))]^2}.
\label{eq:K0}
\end{equation}
The corresponding boundary conditions are 
\begin{equation}
q_0(0) = 0, \quad q_0(\pm \infty) = 1, \quad \tau_0(-\infty) = 0, \quad \tau_0(0) = -\frac{\pi}{2}, \quad \tau_0(\infty) = -\pi.
\end{equation}

\noindent As shown by \cite{amar1991exact} and \cite{tu1991}, the leading-order system \eqref{eq:govO1} can be re-arranged so as to obtain a Ricatti equation,
\begin{equation}
    \frac{\mathrm{d}G}{\mathrm{d}\xi} = G^2(\xi) \frac{\xi}{1+\xi^2} + G(\xi) \left(2\t\frac{\xi}{1+\xi^2} - \frac{1}{\xi}\right) + \t\left(\frac{1}{(1-\lambda)^2} - 1\right)\frac{1+\xi^2}{\xi},
    \label{eq:Ricattiequation}
\end{equation}
\myblue{with boundary conditions,
\begin{equation}
G(\pm \infty) = 0,
\end{equation}} 
\noindent where the new unknown is defined by 
\begin{equation}
G(\xi) = \frac{\e^{\i\tau_0(\xi)}}{q_0(\xi)} - 1.
\end{equation}
\myblue{It was shown by \cite{amar1991exact} and \cite{tu1991} that the} leading-order (zero-surface tension) solutions can then be written in terms of the hypergeometric function $F$, \citep{ASbook},
\begin{subequations} \label{eq:xyhypergeometric}
\begin{align}
x_0 &= (1+\xi^2)^{-\t/2} \times F\left(\frac{\theta_0(2-\lambda)}{2\pi}, - \frac{\lambda \theta_0}{2\pi}, \frac{1}{2}, \frac{\xi^2}{1+\xi^2}\right), \\
 y_0 &= \tilde{A}\, \xi\,(1+\xi^2)^{(1-\t)/2}\times F\left(\frac{1}{2} + \frac{\theta_0 (2-\lambda)}{2\pi}, \frac{1}{2} - \frac{\lambda \theta_0}{2\pi}, \frac{3}{2}, \frac{\xi^2}{1+\xi^2}\right), 
\end{align}
where we have also defined the constant
\begin{equation}
\tilde{A} = 2 \tan \left(\frac{\lambda \theta_0}{2}\right) \cdot\frac{\Gamma\left(1-\frac{\theta_0(2-\lambda)}{2\pi}\right) \Gamma \left(1+ \frac{\lambda\theta_0}{2\pi}\right)}{\Gamma \left(\frac{1}{2} - \frac{\theta_0 (2-\lambda)}{2\pi}\right)\Gamma \left(\frac{1}{2}+\frac{\lambda\theta_0}{2\pi}\right)}.
\end{equation}
\end{subequations}
By differentiating and rearranging (\ref{eq:xyleadingorder}) we also obtain expressions for $q_0(\xi)$ and $\tau_0(\xi)$,
\begin{equation}
q_0 = \t\,\frac{\xi}{1+\xi^2}\,\sqrt{\frac{x_0^2+y_0^2}{\left(\frac{\mathrm{d}x_0}{\mathrm{d}\xi}\right)^2 + \left(\frac{\mathrm{d}y_0}{\mathrm{d}\xi}\right)^2}}, 
\qquad
\cos\tau_0 = -\frac{q_0}{\t}\,\frac{1+\xi^2}{\xi}\,\frac{x_0\frac{\mathrm{d}x_0}{\mathrm{d}\xi} + y_0 \frac{\mathrm{d}y_0}{\mathrm{d}\xi}}{x_0^2 + y_0^2}.
\label{eq:qthhypergeometric}
\end{equation}

\begin{figure}
    \centering
    %\begin{preview}
    %\input{TIKZ/LeadingOrderSolutions}
    %\end{preview}
    \includegraphics{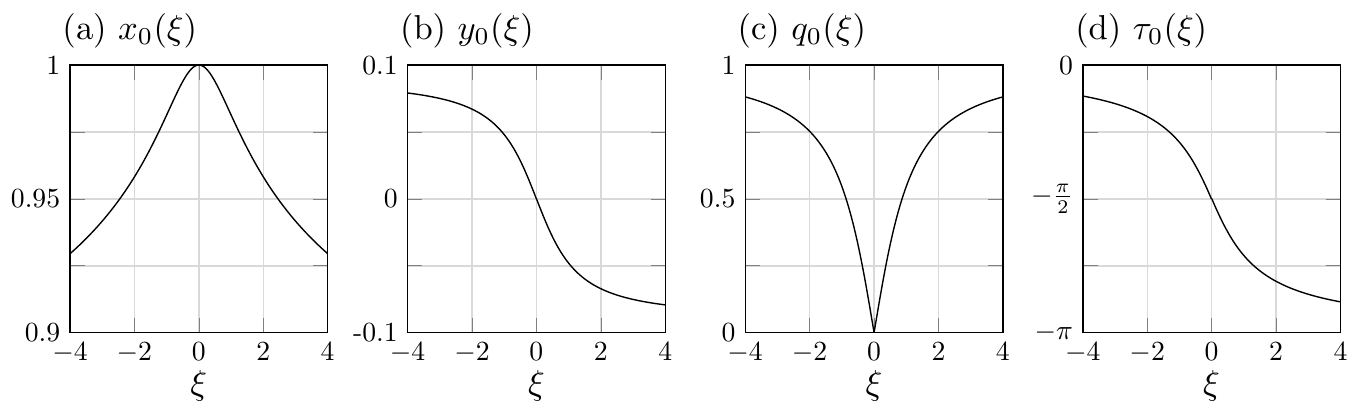}
    \caption{Plots of the leading-order (zero-surface tension) solutions for the four variables $(x_0,y_0,q_0,\tau_0)$ on the free surface, $\zeta = \xi \in \mathbb{Re}$. These plots show the solutions for parameter values $\theta_0 = 20^\circ$ and $\lambda = 0.6$ generated using \eqref{eq:xyhypergeometric} and \eqref{eq:qthhypergeometric}. The tip of the finger lies at $\xi = 0.$ In \cref{fig:HeleShaw} we plot this solution in the physical plane.}
    \label{fig:LeadingOrders}
\end{figure}

\noindent In \cref{fig:LeadingOrders}, we plot example profiles for the leading-order solutions, $(x_0, y_0, q_0, \tau_0)$, along the free surface for parameter values $\theta_0 = 20^\circ$ and $\lambda = 0.6.$ These solutions are generated using \eqref{eq:xyhypergeometric} and \eqref{eq:qthhypergeometric}.

\section{Analytic continuation of the free surface} \label{sec:analyticcontinuationandsingularities}

\noindent The leading-order profiles, as evaluated on the physical free surface, $\zeta = \xi \in \mathbb{R}$, were shown in the previous section. The analytic continuation of these profiles to the complex plane contains square-root singularities. We shall see that such singularities form one of the key ingredients in the exponential asymptotics procedure of \cref{sec:stokeslines}; these points cause the asymptotic expansion to diverge, and will be crucial in determining the eventual selection mechanism. In this section, we discuss the numerical procedure for generating the analytic continuation of the leading-order solutions $(x_0, y_0, q_0, \tau_0)$, as well as the analytic continuation of the governing equations \eqref{eq:bernoulliequation}, \eqref{eq:hilberttransformequation}, \eqref{eq:xyequation}. 

\subsection{Analytic continuation of the leading-order solutions}\label{sec:ancon_leadingorder}

\noindent In the analytic continuation, we allow the previously real-valued $\xi$ to take complex values. \myblue{Keeping in mind the potential for confusion with the previously introduced $\zeta$, we write $\xi = \xi_r + \im \xi_c \mapsto \zeta \in\mathbb{C}$. Note that under this notational choice, $q(\zeta)$ is complex-valued within the fluid region, and it is rather the combination $\Re[q\e^{-\im\tau}/(1-\lambda)]$ that is identified with the fluid speed [cf. \eqref{eq:qtaudef}]}.

In theory, one can replace $\xi$ with $\zeta$, and evaluate  the special-functions solution \eqref{eq:xyhypergeometric} using standard built-in packages (\emph{e.g.} Mathematica) to obtain an analytically-continued leading-order solution. However, the branch structure of the solutions is complicated \myblue{and standard software does not easily allow fine-tune control of the branch placement; generation of the full Riemann surface is subsequently difficult}. In order to develop the results later in the paper, we must implement a scheme that allows for better control over the generation of the Riemann surface and placement of branches.

For this analytic continuation scheme we first split the Ricatti equation \eqref{eq:Ricattiequation} into its real and imaginary parts,
\begin{subequations}\label{eq:leadingorderequationsforcontinuation}
\begin{align}
\frac{\mathrm{d}q_0}{\mathrm{d}\zeta} &= - C(\zeta) \frac{\zeta}{1+\zeta^2}q_0^2 \cos \tau_0 + \frac{q_0}{\zeta} -\t \frac{\zeta}{1+\zeta^2}\cos\tau_0, \\
\frac{\mathrm{d}\tau_0}{\mathrm{d}\zeta} &= -C(\zeta)\frac{\zeta}{1+\zeta^2} q_0 \sin\tau_0 + \frac{\t}{q_0}\frac{\zeta}{1+\zeta^2} \sin\tau_0, 
\end{align}
where
\begin{equation}
C(\zeta) = - \t  + \frac{1}{\zeta^2} \left(1+\zeta^2+ \frac{\theta_0\lambda}{\pi}\left(\frac{2-\lambda}{1-\lambda}\right)\right).
\end{equation}
\end{subequations}
Recall that $\t$ is the rescaled angle parameter introduced in \eqref{eq:elldefinition}. Here, we prefer to use $\zeta$, as we are now working with the complexified version of the Ricatti equation \eqref{eq:Ricattiequation}.

We may now solve the above system along a chosen parameterised path in the complex $\zeta$-plane by using the exact solution on the free surface as an initial condition. \myblue{More specifically, we first pre-solve for $(x_0(\zeta), y_0(\zeta), q_0(\zeta), \tau_0(\zeta))$ on the free surface using \eqref{eq:xyhypergeometric} and \eqref{eq:qthhypergeometric} and setting $\zeta = \xi \in\mathbb{R}$. Then we parameterise a path into the complex plane that starts on the free surface. For example,
\begin{equation}
    \zeta_{path}(s) = \zeta_{IC} + \i s, \quad s\in [0,\infty), \quad \text{where} \quad \zeta_{IC}\in \mathbb{R}.
\end{equation}
We can then solve \eqref{eq:leadingorderequationsforcontinuation} along the parameterised path using any standard ODE integrator (we use MATLAB's ode113 with absolute and relative tolerances set to $10^{-10}$) and the initial condition $\zeta_{IC}$.}

\begin{figure}
    \centering
    \includegraphics{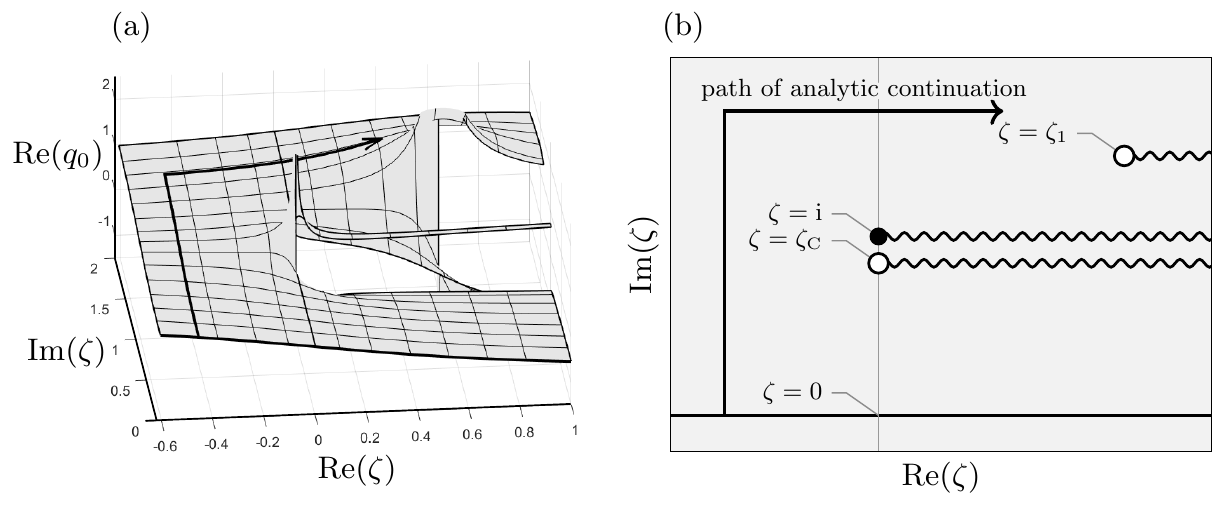}
    \caption{Illustrations of the analytic continuation corresponding to $\theta_0 = 20^{\circ}$ and $\lambda = 0.6$ shown via (a) the $(\Re \zeta, \Im \zeta, \Re q_0)$-plane; and (b) a top-down view of the $\zeta$-plane. In both, a prototypical path of analytic continuation from the physical free-surface is shown with an arrow; branch cuts are shown wavy. 
    Singularities (white circles) lie at $\zeta_1=0.59+1.32\i$ and $\zeta_{\text{C}}=0.96\i$; these both correspond to square root singularities. There is an additional square root branch point at $\i$. The leading-order solution $q_0$ on the free surface lies on the real $\zeta$ axis.}
    \label{fig:analyticcontinuation}
\end{figure}

The solution consists of eight complexified components (the real and imaginary parts of each of the four variables, $(x_0,y_0,q_0,\tau_0)$). In \cref{fig:analyticcontinuation} we show the analytically continued surface for one of these components, $\Re(q_0)$. The figure shows one possible path of analytic continuation. By repeatedly solving along a mesh of different paths, the primary Riemann sheet is generated.

Next we find the location of the singularities in the complex plane numerically. Numerical analytic continuation along a closed loop around a branch point demonstrates that the start- and end-points of the solution differ. Hence continuation around smaller and smaller loops allows the branch point location to be identified. The singularities in this problem are all square-root singularities. Their locations can be found using the method described above; the singularity strength can be further confirmed by verifying the rate of blow-up of the solution as the singularity is approached.

\begin{figure}
    \centering
    \includegraphics{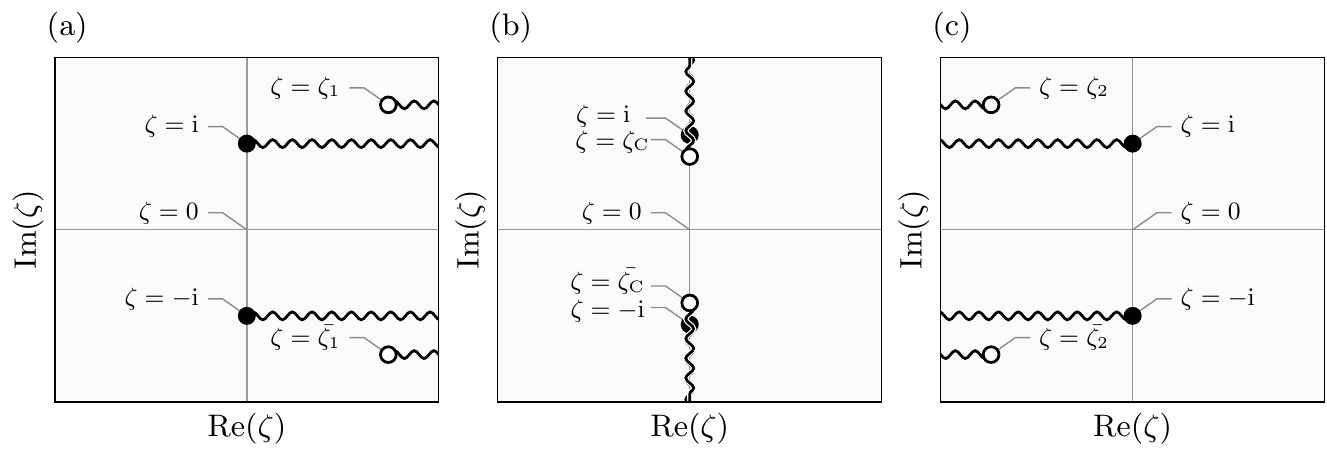}
    %\begin{preview}
    %\input{TIKZ/ACwalk}
    %\end{preview}
    \caption{Plots of the locations of the three complex conjugate pairs of singularities in the $\zeta$-plane: (a) $\{\zeta_1, \bar{\zeta_1}\}$, (b) $\{\zeta_{\text{C}}, \bar{\zeta_{\text{C}}}\}$, (c) $\{\zeta_2, \bar{\zeta_2}\}$. These plots are for parameter values $\theta_0 = 20^\circ$ and $\lambda = 0.6.$ The branch cuts at $\pm \i$ are chosen to show the relevant branches of the Riemann surface which the singularities lie on. The free surface lies on the real $\zeta$ axis, $\zeta = \xi \in \mathbb{R}.$}
    \label{fig:Singularitylocationsketch}
\end{figure}

Using this scheme we find three pairs of complex conjugate singularities, which we denote as $\{\zeta_1, \bar{\zeta_1}\}, \{\zeta_{\text{C}}, \bar{\zeta_{\text{C}}}\}$ and $\{ \zeta_2,\bar{\zeta_2}\}$. The central singularities, $\{\zeta_{\text{C}}, \, \bar{\zeta_{\text{C}}}\}$, lie on the imaginary axis and the non-central singularities $\{\zeta_1, \bar{\zeta_1}\}$ and $\{\zeta_2, \bar{\zeta_2}\}$ lie equidistant from the imaginary axis. The conformal map introduces branch points at $\pm \i$ in the $\zeta$-plane. Consequently, the $\{\zeta_1, \bar{\zeta_1}\}$ singularities do not lie on the same Riemann sheet of the complex $\zeta$-plane as the $\{ \zeta_2, \bar{\zeta_2}\}$ singularities. The three groupings of singularity locations are shown in \cref{fig:Singularitylocationsketch} insets (a)--(c) for the specific case of $\theta_0 = 20^\circ$ and $\lambda=0.6$.

\begin{figure}
    \centering
    %\begin{preview}
    %\input{TIKZ/Singularityzeta1_v2}
    %\end{preview}
    \includegraphics{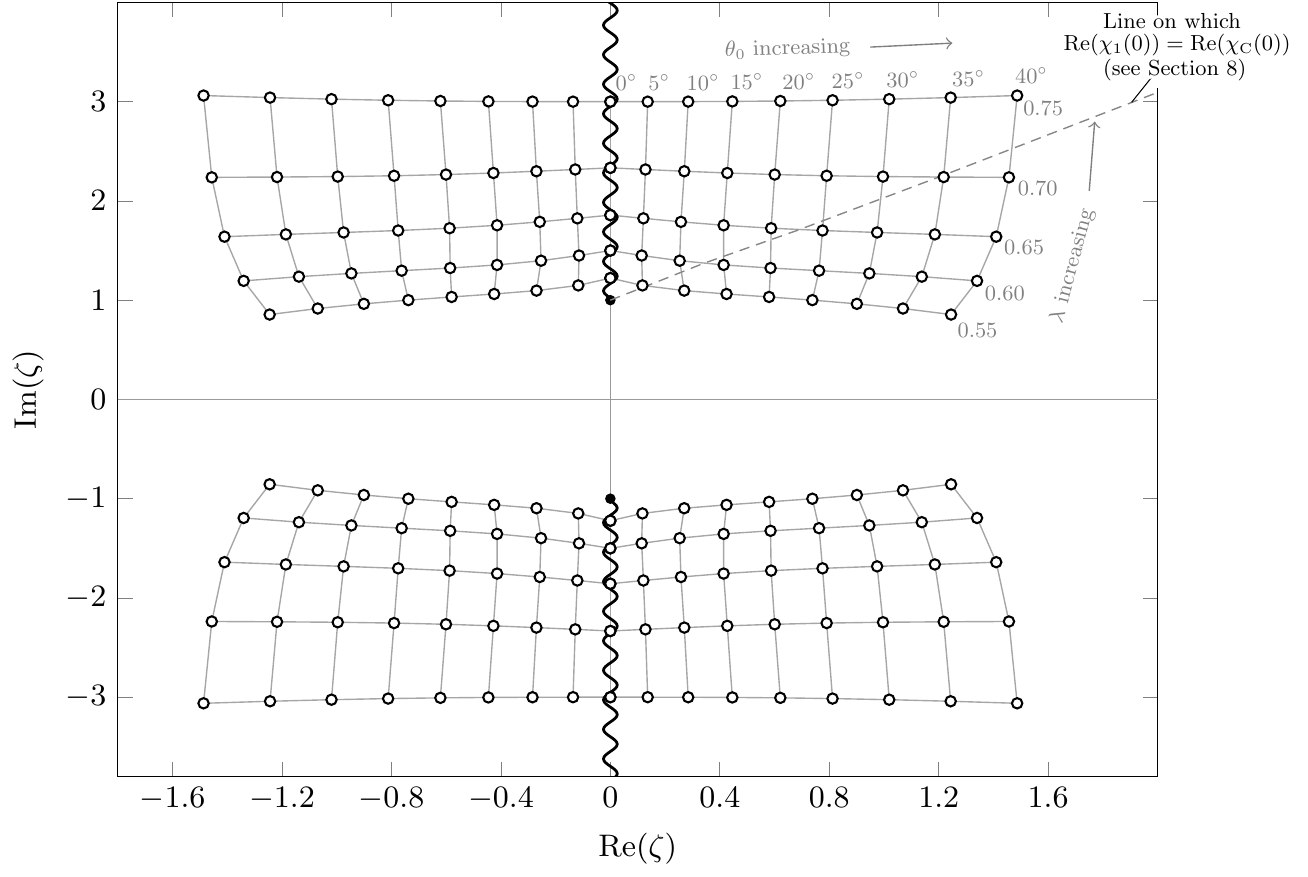}
    \caption{Complex $\zeta$-plane showing how the location of the non-central singularities $\{\zeta_1, \bar{\zeta_1} \}, \{\zeta_2, \bar{\zeta_2} \}$ vary with the parameters $\theta_0$ and $\lambda$. The $\zeta_1$ singularity is shown in the top right quadrant with corresponding $\theta_0$ and $\lambda$ values. The locations of the $\zeta_2,$ $\bar{\zeta_2}$ and $\bar{\zeta_1}$ singularities are shown in the top left, bottom left and bottom right quadrants respectively.}
    \label{fig:zeta1location}
\end{figure}

We can track the locations of the singularities as we vary the parameters $\theta_0$ (the wedge angle) and $\lambda$ (the proportion of the wedge angle occupied by the finger). If $\theta_0>0$ and $\lambda>0.5$, the non-central singularities $\{\zeta_1, \bar{\zeta_1}, \zeta_2, \bar{\zeta_2}\}$ lie off the imaginary axis. In the limit $\theta_0 \rightarrow 0$ the two singularities $\zeta_1$ and $\zeta_2$ converge to the same point on the imaginary axis, but one will be directly above the other on a separate sheet. A reflection of this behaviour occurs for singularities $\bar{\zeta_1}$ and $\bar{\zeta_2}$ in the lower-half $\zeta$-plane. The singularity locations \myblue{in the limit $\theta_0 \to 0$ agree with those found by \cite{chapman1999} in the classic Saffman-Taylor problem with parallel channel walls}. A summary of the locations of these non-central singularities, for varying values of $\theta_0$ and $\lambda$, is shown in \cref{fig:zeta1location}.

Finally, the central singularity, $\zeta_{\text{C}}$, remains on the imaginary axis between the origin and $\i$ for all parameter values. It moves closer to $\i$ as either $\lambda$ decreases towards 0.5, or as $\theta_0$ decreases towards zero. A reflection of this behaviour occurs for $\bar{\zeta_{\text{C}}}$ in the lower half-plane. Example locations of the singularity $\zeta_{\text{C}}$ as $\theta_0$ and $\lambda$ are varied are listed in \cref{table:zetaC}.

\begin{table}
\begin{center}
\begin{tabular}{ c | c c c c c c c c c}
\backslashbox{$\lambda$}{$\theta_0$} & $0^{\circ}$ & $5^{\circ}$ & $10^{\circ}$ & $15^{\circ}$ & $20^{\circ}$ & $25^{\circ}$ & $30^{\circ}$ & $35^{\circ}$ & $40^{\circ}$\\
\hline
 0.55 & $1\i$ & $0.9985\i$ & $0.9660\i$ & $0.9168\i$ & $0.8676\i$ & $0.8205\i$ & $0.7754\i$ & $0.7320\i$ & $0.6900\i$ \\ 
 0.60 & $1\i$ & $1.0000\i$ & $0.9985\i$ & $0.9847\i$ &  $0.9550\i$ & $0.9169\i$ & $0.8755\i$ & $0.8330\i$ & $0.7903\i$\\  
 0.65 & $1\i$ & $1.0000\i$ & $1.0000\i$ & $0.9986\i$ & $0.9913\i$ & $0.9744\i$ & $0.9485\i$ & $0.9164\i$ & $0.8802\i$  \\
 0.70 & $1\i$ & $1.0000\i$ & $1.0000\i$ & $0.9999\i$ & $0.9988\i$ & $0.9945\i$ & $0.9845\i$ & $0.9680\i$ & $0.9451\i$\\
 0.75 & $1\i$ & $1.0000\i$ & $1.0000\i$ & $1.0000\i$ & $0.9999\i$ & $0.9991\i$ & $0.9964\i$ & $0.9905\i$ & $0.9802\i$
\end{tabular}
\end{center}
\caption{The locations of the central singularity, $\zeta_{\text{C}}$ (to 4 s.f.) for different values of $\theta_0$ and $\lambda$.}
\label{table:zetaC}
\end{table}

\subsection{Analytic continuation for the full problem}

\noindent If the variable $\zeta$ is analytically continued into the upper-half $\zeta$-plane, the Bernoulli equation \eqref{eq:bernoulliequation} becomes
\begin{subequations} \label{eq:govAC}
\begin{equation}
\epsilon^2 \frac{P(\zeta)}{4} \pd{}{\zeta} \left(r(x) \left[P(\zeta) q(\zeta) \pd{\tau}{\zeta} - \t \sin\tau(\zeta)\right]\right) 
= \frac{2}{\pi} \int_{-\infty}^0 \frac{K(\tilde{\zeta}) \,\tilde{\zeta}}{\zeta^2-\tilde{\zeta}^2}\de{\tilde{\zeta}} + \i K(\zeta).
\label{eq:bernoulliAC}
\end{equation}
Here the principle value integral becomes a normal integral and can then be combined with the $Q_0$ term in \eqref{eq:bernoulliequation} to simplify the right hand-side. Recall that $P(\zeta), r(x(\zeta)),\t, K(\zeta)$ were introduced in \eqref{eq:PrK} for convenience.

To analytically continue the boundary integral equation \eqref{eq:hilberttransformequation} we must consider the complexification of the Hilbert transform,
\begin{equation}
    \mathcal{H}[\tau] = \hat{\mathcal{H}}[\tau] + \i \tau,
\end{equation}
where $\hat{\mathcal{H}}[\tau]$ is the complex-valued Hilbert transform,
\begin{equation}
    \hat{\mathcal{H}}[\tau](\xi) = \frac{2}{\pi}\int_{-\infty}^0 \frac{\tau(\tilde{\xi}) \, \tilde{\xi}}{\xi^2 - \tilde{\xi}^2} \, \de{\tilde{\xi}}.
\end{equation}
The boundary integral equation \eqref{eq:hilberttransformequation} then becomes
\begin{equation}
\log q(\zeta) - \i \tau(\zeta) = \hat{\mathcal{H}}[\tau](\zeta),
\label{eq:hilbertransformequationAC}
\end{equation}
and hence the integrated equation for the surface position (\ref{eq:xyequation}) becomes, 
\begin{equation}
x(\zeta)+\i y(\zeta) = \frac{2(1-\lambda)}{\pi}\int_\zeta^0 \frac{\exp\left(-\hat{\mathcal{H}}[\tau](\tilde{\zeta})\right)}{P(\tilde{\zeta})}\de{\tilde{\zeta}}.
\label{eq:xyequationAC}
\end{equation}
\end{subequations}
This results in a set of analytically continued governing equations \eqref{eq:govAC} that hold in the upper-half $\zeta$-plane.

\section{Exponential asymptotics}

\noindent Our procedure for the exponential asymptotic analysis follows similar work by \cite{Tanveer1987} and \cite{chapman1999}, using the methodology established in \cite{Chapman1998}. In essence, \myblue{the goal is to derive the behaviour of the late terms in the asymptotic series.} After, in \cref{sec:stokeslines}, these late terms are used to study the exponentially-small terms via the Stokes-line switching.  

As $\epsilon \rightarrow 0$, we expand the dependent variables as 
\begin{equation}
\begin{gathered}
    x(\zeta) \sim \sum_{n=0}^\infty \epsilon^{2n}x_n(\zeta), \qquad y(\zeta) \sim \sum_{n=0}^\infty \epsilon^{2n}y_n(\zeta), \\
    q(\zeta) \sim \sum_{n=0}^\infty \epsilon^{2n}q_n(\zeta), \qquad \tau(\zeta) \sim \sum_{n=0}^\infty \epsilon^{2n}\tau_n(\zeta). \label{eq:qtaupowerseries}
\end{gathered}    
\end{equation}
We substitute the above into the analytically continued governing equations \eqref{eq:govAC} and this yields, at $\mathcal{O}\left(\epsilon^{2n}\right)$ for Bernoulli's equation \eqref{eq:bernoulliAC},
\begin{subequations}\label{eq:ordernequations}
\begin{equation}
\begin{aligned}
\frac{P}{4}\frac{\partial}{\partial \zeta}\Bigg(r(x_0)\Bigg[Pq_{n-1} \frac{\partial \tau_0}{\partial \zeta} +& Pq_0 \frac{\partial \tau_{n-1}}{\partial \zeta}  - \t \tau_{n-1}\cos\tau_0 \\
-\frac{\theta_0 x_{n-1}}{2} &\left(Pq_0 \frac{\partial \tau_0}{\partial \zeta} - \t \sin\tau_0 \right) + \ldots \Bigg] \Bigg)  \\
&= \ldots +  \frac{\i}{r^2(x_0)\,q_0}\left(\tau_n\cos\tau_0 - \frac{q_n}{q_0}\sin\tau_0 +\ldots\right),
\label{eq:bernoullinthorderlong}
\end{aligned}
\end{equation}
while for \eqref{eq:hilbertransformequationAC} and \eqref{eq:xyequationAC} we have
\begin{align}
\frac{q_n}{q_0} + \ldots - \i\tau_n &= \hat{\mathcal{H}}[\tau_n], \label{eq:hilbertransformnthorder} \\
x_n + \i y_n &= \frac{2(1-\lambda)}{\pi}\int_\zeta^0 \frac{\exp(-\hat{\mathcal{H}}[\tau_0]) \hat{\mathcal{H}}[\tau_n] + \ldots}{P}\de{\tilde{\zeta}}. \label{eq:xynthorder}
\end{align}
\end{subequations}
At these later orders we see that the $n$\textsuperscript{th} terms in the power series for $q$ and $\tau$ are obtained by differentiating the $(n-1)$\textsuperscript{th} terms twice. Any singularities in the leading-order solution will grow in strength with each successive differentiation. This means that later terms in the power series will have singularities in the same locations as earlier terms, but with increasing strength. We therefore follow the method of \cite{Chapman1998} and predict a factorial-over-power form for the late-order terms, 
\begin{equation}
q_n(\zeta) \sim \frac{Q(\zeta) \,\Gamma (2n+\gamma)}{\chi(\zeta)^{2n+\gamma}}, \quad \tau_n(\zeta) \sim \frac{\Theta(\zeta)\,\Gamma(2n+\gamma)}{\chi(\zeta)^{2n+\gamma}}, \quad \text{as} \quad n \rightarrow \infty
\label{eq:factorialoverpower}
\end{equation} 
where $Q$ and $\Theta$ are prefactors and $\chi$ is a singulant function, which is zero at the singularity. The singulant ensures that each series term has singularities with the correct locations and $\gamma$ ensures they have the correct strengths. The Gamma function \citep{ASbook} is a consequence of the factorial behaviour caused by repeated differentiation. The late-order terms are a sum of such factorial-over-power terms -- one associated with each distinct complex singularity. \myblue{Note that the prototypical factorial-over-power divergence of singular asymptotic expansions is a consequence of Darboux's theorem [cf. p.~4 of \citealt{Dinglebook} and \cite{Crew2023}]}.

In the limit $n\to\infty$, the behaviour of the asymptotic expansion will be dictated by the divergence caused by the singularities driving \eqref{eq:factorialoverpower} \citep{Chapman1998}. From \cref{sec:analyticcontinuationandsingularities} we know that the singularities lie in the complex plane away from the free surface. The complex Hilbert transform, $\hat{\mathcal{H}}[\tau_n]$, involves the integrand evaluated along the free surface. \myblue{Once the ansatzes for $q_n$ and $\tau_n$ via \eqref{eq:factorialoverpower} are substituted into the Hilbert transform, we may observe that the contribution of $\hat{\mathcal{H}}[\tau_n]$ will be subdominant in the limit that $n \rightarrow \infty$, compared to $q_n$ and $\tau_n$. This follows from the increasing nature of $|\chi|$ along Stokes lines, as explained on p.~526 of \cite{chapman1999}}. The combination of $x_n+\i y_n$ will also be subdominant in this limit, although the individual components may still diverge. We will assume in this analysis that the individual components do not diverge, and this assumption will be validated \textit{a posteriori}, as is done in similar asymptotic analyses of boundary-integral problems~\citep{Shelton2022}.

Using these assumptions gives the dominant behaviour from the boundary integral equation \eqref{eq:hilbertransformnthorder}
\begin{equation}
\frac{q_n}{q_0} - \frac{q_1q_{n-1}}{q_0^2} \sim \i\tau_n  \quad \text{as} \quad n \rightarrow \infty.
\label{eq:hilbertransformnthorder2}
\end{equation}
The Bernoulli equation \eqref{eq:bernoullinthorderlong} can then be simplified to
\begin{equation}
\frac{q_{n-1}''}{q_n} + \left[\frac{\left(r(x_0)P\right)'}{r(x_0)P} - \frac{q_0'}{q_0} + \i\tau_0' - \t \frac{\cos\tau_0}{Pq_0}\right]\frac{q_{n-1}'}{q_n} = \frac{4(\sin\tau_0 + \i \cos\tau_0)}{P^2q_0^2r(x_0)^3}.
\label{eq:bernoullinthorder}
\end{equation}
Here, and for the rest of the paper, we use primes ($'$) to denote differentiation with respect to $\zeta$. Substitution of  the factorial-over-power ansatz (\ref{eq:factorialoverpower}), and matching terms in the limit that $n \rightarrow \infty$ gives at leading-order an equation for the singulant, $\chi$:
\begin{equation}
(\chi')^2 = \frac{4(\sin\tau_0 + \i\cos\tau_0)}{P^2q_0^2r(x_0)^3},
\label{eq:singulant}
\end{equation}
which can be solved to obtain
\begin{equation}
\chi = -\int_{\zeta_*}^\zeta \frac{2\sqrt{\sin\tau_0 + \i\cos\tau_0}}{P\,q_0\,r(x_0)^{\frac{3}{2}}} \de{\tilde{\zeta}}.
\label{eq:singulantintegrated}
\end{equation}
In this expression, $\zeta_*$ is the singularity location and the integration limits are chosen so that $\chi(\zeta_*) = 0$. There will be a singulant function $\chi$ for each complex singularity. The negative sign is selected when taking the square root so that the Stokes line intersects the free surface, which lies on the real $\zeta$ axis. 

Matching terms in (\ref{eq:bernoullinthorder}) at the next order as $n \rightarrow \infty$ shows that $\gamma$ is constant, and at the following order we obtain,
\begin{equation}
\frac{Q'}{Q} = - \frac{1}{2}\left[\frac{\chi''}{\chi'} + \frac{(r(x_0)P)'}{r(x_0)P} - \frac{q_0'}{q_0} + \i\tau_0' - \t \frac{\cos\tau_0}{Pq_0}\right],
\end{equation}
which can be solved to give an expression for the prefactor,
\begin{equation}
\quad Q = \Lambda \i \left(\frac{q_0\, \e^{-\i\tau_0} \exp \left[\int_0^\zeta \t \frac{\cos\tau_0}{Pq_0} \de{\tilde{\zeta}}\right]}{\chi' \,r(x_0)\,P}\right)^{\frac{1}{2}},
\label{eq:prefactorQ}
\end{equation}
where $\Lambda$ is a constant that remains to be determined. From the boundary integral equation \eqref{eq:hilbertransformnthorder2} we find,
\begin{equation}
\Theta = -\i\frac{Q}{q_0}.
\label{eq:Prefactorrelationship}
\end{equation}
\myblue{By definition, $\ell \rightarrow 0$ (see \eqref{eq:elldefinition}) and $r(x_0) \rightarrow 1$ (see \eqref{eq:PrK}) in the limit $\theta_0 \rightarrow 0$. This means that these results are consistent with those found by \cite{chapman1999} for the Saffman-Taylor problem in a channel geometry.}

The late-order series terms in the divergent power series solution for $q(\zeta)$ therefore have the form,
\begin{equation}
q_n \sim \Lambda \i \left(\frac{q_0 \e^{-\i\tau_0} \exp \left[\int_0^\zeta \t \frac{\cos\tau_0}{Pq_0} \de{\tilde{\zeta}}\right]}{\chi' r(x_0)P}\right)^{\frac{1}{2}} \frac{\Gamma(2n+ \gamma)}{\chi^{2n+\gamma}}, \quad \text{as} \quad n\rightarrow \infty.
\label{eq:lateorderq}
\end{equation}

\section{Optimal truncation and Stokes lines} \label{sec:stokeslines}

\noindent In the previous section we noted that complex singularities cause the power series to become divergent and so they will need to be truncated. We will truncate the divergent power series at some to be determined optimal point $N$, and introduce the remainder terms, 
\begin{subequations}
\begin{align}
x = \sum_{n=0}^{N-1} \epsilon^{2n} x_n + R_x, &\quad y = \sum_{n=0}^{N-1} \epsilon^{2n}y_n + R_y, \\
q  = \sum_{n=0}^{N-1} \epsilon^{2n} q_n + R_q, &\quad \tau = \sum_{n=0}^{N-1} \epsilon^{2n} \tau_n + R_\tau.
\end{align}
\end{subequations}
We can substitute these into the governing equations \eqref{eq:govAC}. Given that the first $N$ orders must exactly satisfy the relationships in \eqref{eq:ordernequations}, we derive leading-order relationships between the remainder terms.

From the boundary integral equation (\ref{eq:hilbertransformnthorder2}) we derive at leading-order as $\epsilon\rightarrow 0$ that 
\begin{equation}
\frac{R_q}{q_0} \sim \i R_\tau.
\end{equation}
Using this in the $x+\i y$ equation \eqref{eq:xynthorder} we see that $R_x$ and $R_y$ are subdominant compared to the leading-orders of $R_q$ and $R_\tau$ in the limit $\epsilon \rightarrow 0$.

We substitute these relationships into the Bernoulli equation \eqref{eq:bernoulliAC} and derive a single ordinary differential equation for $R_q$, which we now rename as $R_N$ for consistency with other works including  \cite{chapman1999}. This ordinary differential equation, known as the remainder equation, reduces to,\myblue{
\begin{equation}
\epsilon^2R_N'' + \epsilon^2 \left[\frac{(r(x_0)P)'}{r(x_0)P} -\frac{q_0'}{q_0}+ \i \tau_0' - \t  \frac{1}{P q_0}\cos\tau_0\right]R_N' = (\chi')^2(R_N-\epsilon^{2N}q_N),
\end{equation}
as $\epsilon \rightarrow 0$, where the terms that do not appear at the leading or second-order in this limit have been omitted}. Changing the independent variable to $\chi$ (primes will continue to denote differentiation with respect to $\zeta$) gives,
\begin{equation}
\epsilon^2 \frac{\mathrm{d}^2R_N}{\mathrm{d}\chi^2} + \epsilon^2 \frac{1}{\chi'} \left[\frac{\chi''}{\chi'} + \frac{(r(x_0)P)'}{r(x_0)P} -\frac{q_0'}{q_0}+ \i \tau_0' - \t  \frac{1}{P q_0}\cos\tau_0\right]\frac{\mathrm{d} R_N}{\mathrm{d}\chi} - R_N = - \epsilon^{2N}q_N.
\end{equation}
Using the definition for the prefactor (\ref{eq:prefactorQ}), this can be simplified to,
\begin{equation}
\epsilon^2 \frac{\mathrm{d}^2R_N}{\mathrm{d}\chi^2} -2\epsilon^2\frac{Q'}{Q\chi'}\frac{\mathrm{d} R_N}{\mathrm{d}\chi} - R_N = -\epsilon^{2N}q_N.
\label{eq:remainderequation}
\end{equation}

\noindent To solve \eqref{eq:remainderequation} we pose a Liouville-Green or WKBJ-style ansatz for the form of the remainder given by
\begin{equation}
R_N = B(\chi)e^{\frac{b(\chi)}{\epsilon}}.
\label{eq:WKBremainder}
\end{equation}
Then equating the coefficients at different powers of $\epsilon$ for the homogeneous version of the remainder equation (\ref{eq:remainderequation}), we find that $b = \pm \chi + \text{constant}$ and $B\sim Q.$ The arbitrary constant in $b$ is equivalent to multiplying the entire expression \eqref{eq:WKBremainder} by an arbitrary constant that is not determined by the WKBJ analysis.

To solve the full inhomogeneous remainder equation (\ref{eq:remainderequation}) we apply the method of variation of parameters and permit the arbitrary constant to vary in $\zeta$. This quantity is known as the Stokes multiplier, and we denote it by $A(\chi)$. The remainder becomes, 
\begin{equation}
R_N = A Q \e^{-\frac{\chi}{\epsilon}},
\end{equation}
where the negative sign in the exponent ensures the remainder is exponentially small.
The inhomogeneous equation (\ref{eq:remainderequation}) gives
\begin{equation}
-2\epsilon \frac{\mathrm{d}A}{\mathrm{d}\chi} Q \e^{-\frac{\chi}{\epsilon}} = -\epsilon^{2N} \frac{Q\Gamma(2N+\gamma)}{\chi^{2N+\gamma}},
\label{eq:StokesmultiplierODE}
\end{equation}
where we have substituted in the late-order expression for $q_N$ from \eqref{eq:factorialoverpower}.

The next step involves truncating the series at an optimal point. We define this optimal truncation point, $N$, to be where successive terms in the divergent series are approximately equal in magnitude, so
\begin{equation}
\left| \frac{\epsilon^{2N+2} q_{N+1}}{\epsilon^{2N}q_N}\right| \approx 1.
\end{equation}
This condition gives $N \approx \frac{\left|\chi\right|}{2\epsilon}$. As $N$ must be an integer we let $N = \frac{|\chi|}{2\epsilon} +\beta,$ where $\beta$ is bounded as $\epsilon\rightarrow 0.$ This form motivates the transformation to polar coodinates, so we define $\chi = |\chi|\e^{\i\eta}.$

\begin{figure}
    \centering
    %\begin{preview}
    %\input{TIKZ/StokesLinesSheets}
    %\end{preview}
    \includegraphics{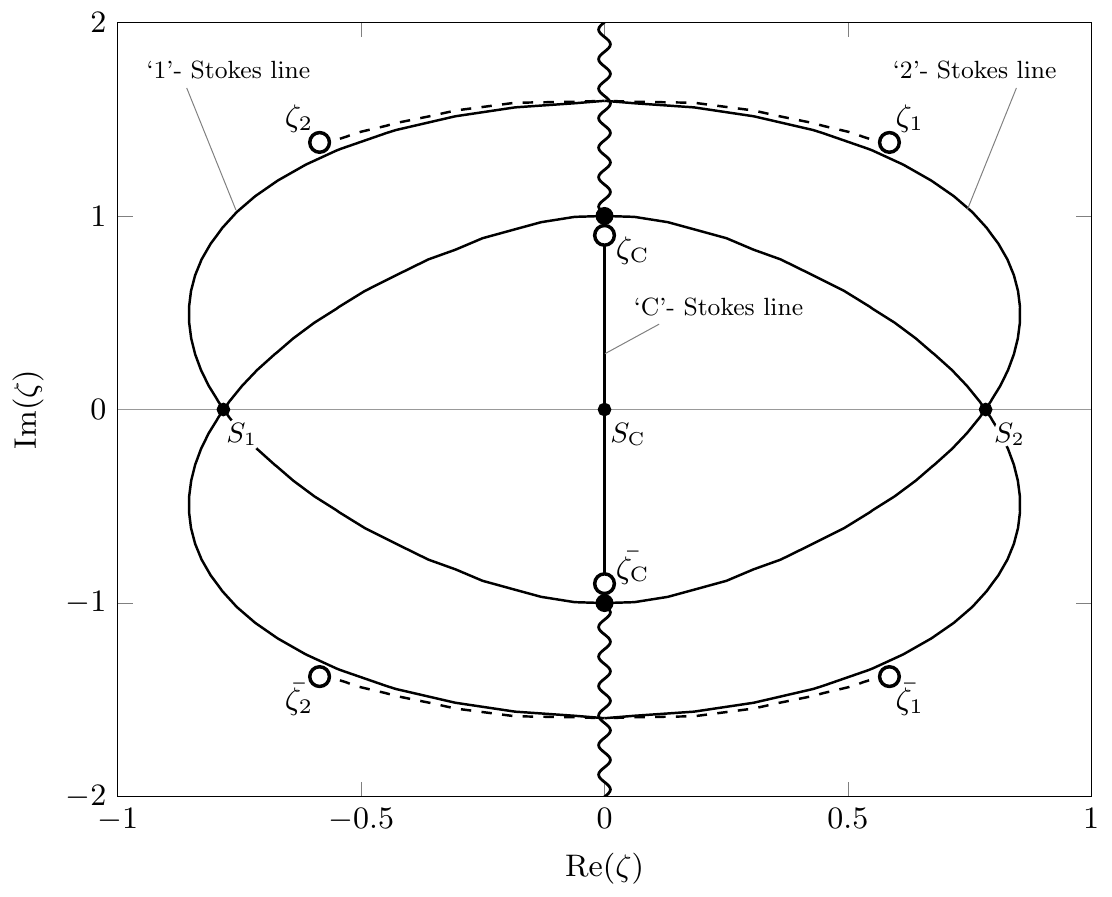}
    \caption{Complex $\zeta$-plane showing the Stokes lines emanating from the singularities (circles) and intersecting the free surface (real axis). In this figure, the wedge angle is $\theta_0 = 20^\circ$ and $\lambda = 0.6$. Branch cuts (shown wavy) lie up and down the imaginary axis from $\pm \i$. Stokes lines are shown with dashed when they lie on a different Riemann sheet to the free surface. The three points where the Stokes lines intersect the real axis are labelled $S_1,$ $S_{\text{C}}$ and $S_2$ respectively.}
    \label{fig:Stokeslines}
\end{figure}

By the chain rule we have
\begin{equation}
    \frac{\mathrm{d}A}{\mathrm{d}\eta} = \frac{\mathrm{d}A}{\mathrm{d}\chi}\i\chi.
    \label{eq:stokesmultiplierchainrule}
\end{equation}
We substitute the optimal value of $N$ into \eqref{eq:StokesmultiplierODE} and note that $N$ is large which allows us to use Stirling's formula \citep{ASbook} to approximate $N!$ in the large $N$ limit. Then using \eqref{eq:stokesmultiplierchainrule} we obtain,
\begin{equation}
\frac{\mathrm{d}A}{\mathrm{d}\eta} \sim \frac{\i \epsilon^{2N-1}}{2}\frac{\exp\left(\frac{|\chi|}{\epsilon}\e^{\i\eta}\right) \sqrt{2\pi} \left(2N+\gamma\right)^{2N+\gamma-\frac{1}{2}}}{\e^{2N+\gamma}|\chi|^{2N+\gamma-1}\e^{\i\eta(2N+\gamma-1)}}.
\end{equation}
We substitute in the optimal value of $N$ to give,
\begin{equation}
\frac{\mathrm{d}A}{\mathrm{d}\eta} \sim \frac{\i\sqrt{\pi}}{\sqrt{2}c}\frac{|\chi|^{\frac{1}{2}}}{\epsilon^{\gamma+\frac{1}{2}}}\frac{\e^{2\beta+\gamma}}{\exp\left(\i\eta\left( \frac{|\chi|}{\epsilon}+2\beta+\gamma\right)\right)}\frac{\exp\left(\frac{|\chi|}{\epsilon}\e^{\i\eta}\right)}{\exp\left(\frac{|\chi|}{\epsilon} + 2\beta+\gamma\right)} \quad \text{as} \quad \epsilon \rightarrow 0.
\label{eq:StokesmultiplierODEinner}
\end{equation}
The right hand side of \eqref{eq:StokesmultiplierODEinner} is exponentially small in $\epsilon$ unless $\eta = 2\pi k,$ $ k \in \mathbb{Z},$ in which case it will be algebraic in the limit that $\epsilon\rightarrow 0.$ Therefore, the greatest change in the Stokes multiplier $A$ will occur across curves, or Stokes lines, on which $\eta = 2\pi k$, and so
\begin{equation} \label{eq:stokeslinedefinition}
    \mathrm{Re}(\chi)>0, \quad \mathrm{Im}(\chi) = 0.
\end{equation}
\myblue{This recovers the classic result of \cite{Dinglebook}.} We compute the Stokes lines numerically, with the results shown in \cref{fig:Stokeslines}. 

When Stokes lines intersect the free surface (the real axis in the $\zeta$-plane) then an exponentially small term will be smoothly switched on across this intersection point in the solution. We can see from \cref{fig:Stokeslines} that there are three points on the free surface where such exponentially small terms will be switched on.

To find the jump in solution behaviour across a Stokes line, we rescale $\eta$ and $A$ and consider the behaviour in the neighbourhood of the Stokes line. We apply the rescaling,
\begin{equation}
\eta = \epsilon^2 \tilde{\eta} \quad \text{and} \quad A = \tilde{A}\epsilon^{-\gamma-1}.
\end{equation}
We let $k = 0$ and then \eqref{eq:StokesmultiplierODEinner} becomes
\begin{equation}
\frac{\mathrm{d}\tilde{A}}{\mathrm{d}\tilde{\eta}} \sim \frac{\i\sqrt{\pi}}{\sqrt{2}} |\chi|^{\frac{1}{2}}\exp\left(-\frac{1}{2}|\chi| \tilde{\eta}^2\right) \quad \text{as} \quad \epsilon\rightarrow 0.
\end{equation}
Integrating this gives
\begin{equation}
\tilde{A} = \text{constant} + \i\sqrt{\pi} \int_{-\infty}^{\tilde{\eta}\sqrt{\frac{|\chi|}{2}}} \e^{-\tilde{\xi}^2}\de{\tilde{\xi}},
\label{eq:stokesconstantintegral}
\end{equation}
and hence the jump in the Stokes multiplier across the Stokes line is,
\begin{equation}
\lim_{\eta \rightarrow 0+}A(\eta) - \lim_{\eta \rightarrow 0-}A(\eta) = \frac{\i\sqrt{\pi}}{\epsilon^{\gamma+1}} \int_{-\infty}^\infty \e^{-\tilde{\xi}^2}\de{\tilde{\xi}} = \frac{\i\pi}{ \epsilon^{\gamma+1}}.
\label{eq:stokesmultiplierjump}
\end{equation}
That is, upon crossing a Stokes line at its intersection with the real axis, a contribution in the $q$ variable is switched on. This has the form,
\begin{equation}
    \frac{\i \pi}{  \epsilon^{\gamma+1}}Q\exp\left(-\frac{\chi}{\epsilon}\right) + \text{c.c.},
\end{equation}
where c.c. represents the complex conjugate expression, which arises from the singularity located at the complex conjugate location in the complex plane (\cref{fig:Stokeslines}). Similarly, by (\ref{eq:Prefactorrelationship}), the contribution that is switched on in the $\tau$ variable has the form,
\begin{equation}
    \frac{\pi}{ \epsilon^{\gamma+1}} \frac{Q}{q_0}\exp\left(-\frac{\chi}{\epsilon}\right) + \text{c.c.}.
    \label{eq:StokeslineJump}
\end{equation}
In the next section we show how these terms switched on across Stokes lines will determine the selection mechanism.

\section{Selection mechanism}

\noindent In \cref{fig:Stokeslines} we see that there are three points on the free surface ($\zeta = \xi \in \mathbb{R}$) that are intersected by Stokes lines. We will continue the notation from the singularities \myblue{(introduced in \cref{sec:ancon_leadingorder})} and use the subscripts `$1$', `$\text{C}$' and `$2$' to label the Stokes lines, \emph{i.e.} each subscript matches the respective label of the associated Stokes-line singularity. We denote the three intersection points as $S_1$, $S_{\mathrm{C}}$ and $S_2$.  Each Stokes line will have different corresponding values for $\Lambda,$ $\gamma$ and $\chi$, which will also be labelled with the same subscripts. 

At each intersection point an exponentially small asymptotic contribution of the form (\ref{eq:StokeslineJump}) will be switched on or off. The far-field conditions $\tau(-\infty) = 0$ and $\tau( \infty) = -\pi$ imply there are no exponentially small contributions present on the free surface as $\tau \to \pm\infty$, so the exponential terms must only be present in the region of the free surface between the Stokes line intersection points; that is, in the range $\zeta=\xi\in (S_1, S_2)$. 

To check for existence of a solution we can check that the conditions in both farfields are satisfied. However, due to the symmetry of the problem it is equivalent to consider half the domain and impose symmetry conditions at the origin: $q(0) = 0,$ $\tau(0) = -\pi/2$. A similar symmetry condition is used in \cite{Chapman2023}. Enforcing the condition on $\tau$ implies,
\begin{equation}\label{eq:origincond}
    \tau_0(0) = -\frac{\pi}{2},\qquad \tau_n(0) = 0 \quad \mathrm{for} \quad n\geq 1, \qquad R_\tau(0) = 0.
    \end{equation}
Requiring that the solution satisfies the farfield conditions as $\xi \to -\infty$ and the conditions at the origin from (\ref{eq:origincond}) will result in a selection condition that must be satisfied for solutions to exist.

At the point $S_1$, a contribution of the form (\ref{eq:StokeslineJump}) is switched on as the Stokes line is crossed from left to right. And then to reach the origin we switch on half of another contribution of the form (\ref{eq:StokeslineJump}) (as we have only crossed half the boundary layer about the `$\text{C}$'-Stokes line). That is, the integral in \eqref{eq:stokesconstantintegral} will only range from $-\infty$ to $\tilde{\eta} = 0.$ The symmetry condition from (\ref{eq:origincond}) says that the sum of these contributions must be zero at the origin. This means,
\begin{equation}
    \begin{aligned}
        \pi \epsilon^{-\gamma_1-1}&\frac{\Lambda_1}{\sqrt{2}} \e^{-\i{\tau_0}/{4}}I(\zeta)r(x_0)^{\frac{1}{4}}\e^{3\i{\pi}/{8}} \exp\left(-\frac{\chi_1}{\epsilon}\right) + \mathrm{c. c.} \\
        +\frac{1}{2}\pi& \epsilon^{-\gamma_{\text{C}}-1}\frac{\Lambda_C}{\sqrt{2}} \e^{-\i{\tau_0}/{4}}I(\zeta)r(x_0)^{\frac{1}{4}}\e^{3\i{\pi}/{8}} \exp\left(-\frac{\chi_{\text{C}}}{\epsilon}\right) + \mathrm{c. c.}\Bigg|_{\zeta=0} = 0,
    \end{aligned}
\end{equation}
where $I(\zeta) := \exp\left[\int_0^\zeta \t \frac{\cos\tau_0}{2Pq_0}\de{\tilde{\zeta}}\right].$
We simplify this expression using the conditions at the origin, $\tau_0(0) = -{\pi}/{2}$ and $x_0(0) = 0$. Noting that the origin lies on the `$\text{C}$' Stokes line, we use the definition for a Stokes line obtained in \eqref{eq:stokeslinedefinition}, \emph{i.e.} $\mathrm{Im}(\chi_{\text{C}}(0)) = 0$. Then,
\begin{equation}
    \begin{aligned}
        \frac{|\Lambda_1|}{|\Lambda_{\text{C}}|} \epsilon^{-\gamma_1+\gamma_{\text{C}}}\e^{-\mathrm{Re}(\chi_1(0))/\epsilon+\mathrm{Re}(\chi_{\text{C}}(0))/\epsilon} \cos\Bigg(\arg(\Lambda_1) +& \frac{\pi}{2} - \frac{\mathrm{Im}(\chi_1(0))}{\epsilon}\Bigg) \\
        &+ \frac{1}{2}\cos\left(\arg(\Lambda_{\text{C}}) + \frac{\pi}{2}\right) = 0.
    \end{aligned}
\end{equation}
By the symmetry of the problem this condition could also be obtained by calculating the values that appear across the `$2$' and `$\text{C}$' Stokes lines. Note that for this problem we always have symmetry of $q_0$ at the origin and so $q_0(0) = q_0'(0) = 0$. This means that the possible selection condition, $q'(0) = 0$, is automatically satisfied. We therefore use the $\tau$ variable to derive the selection mechanism. 

The constants $\gamma_1$ and $\gamma_{\text{C}}$ are derived by an inner matching of the local singularity strengths at lower orders in the asymptotic series. The details of the derivation can be found in Appendix \ref{app:gamma}. We find that $\gamma_1 = \gamma_{\text{C}} = {1}/{14}$ and thus we can simplify the selection condition to
\begin{multline}
\frac{|\Lambda_1|}{|\Lambda_{\text{C}}|} \e^{-\mathrm{Re}(\chi_1(0))/\epsilon+\mathrm{Re}(\chi_{\text{C}}(0))/\epsilon} \cos\left(\arg(\Lambda_1) + \frac{\pi}{2} - \frac{\mathrm{Im}(\chi_1(0))}{\epsilon}\right) \\
+ \frac{1}{2}\cos\left(\arg(\Lambda_{\text{C}}) + \frac{\pi}{2}\right) = 0.
\label{eq:selectioncondition}
\end{multline}
Above, the constants $\Lambda_1$ and $\Lambda_{\text{C}}$ are derived in Appendix \ref{app:lambda} by using an asymptotic matching process to connect the late-order term behaviour (\ref{eq:factorialoverpower}) with a local expansion of the solution in a neighbourhood of the relevant singularity. We perform this matching numerically, and find that the values of $\Lambda_1$ and $\Lambda_{\text{C}}$ depend on both the parameters $\theta_0$ and $\lambda$. For example, when $\theta_0 = 20^\circ$ and $\lambda = 0.6$ then $\Lambda_1 \approx -0.48-0.15\i$ and $\Lambda_{\text{C}} \approx -0.49-0.24\i$.

\section{Results}

\begin{figure}
    \centering
    %\begin{preview}
    %\input{TIKZ/BifurcationDiagrams}
    %\end{preview}
        \includegraphics{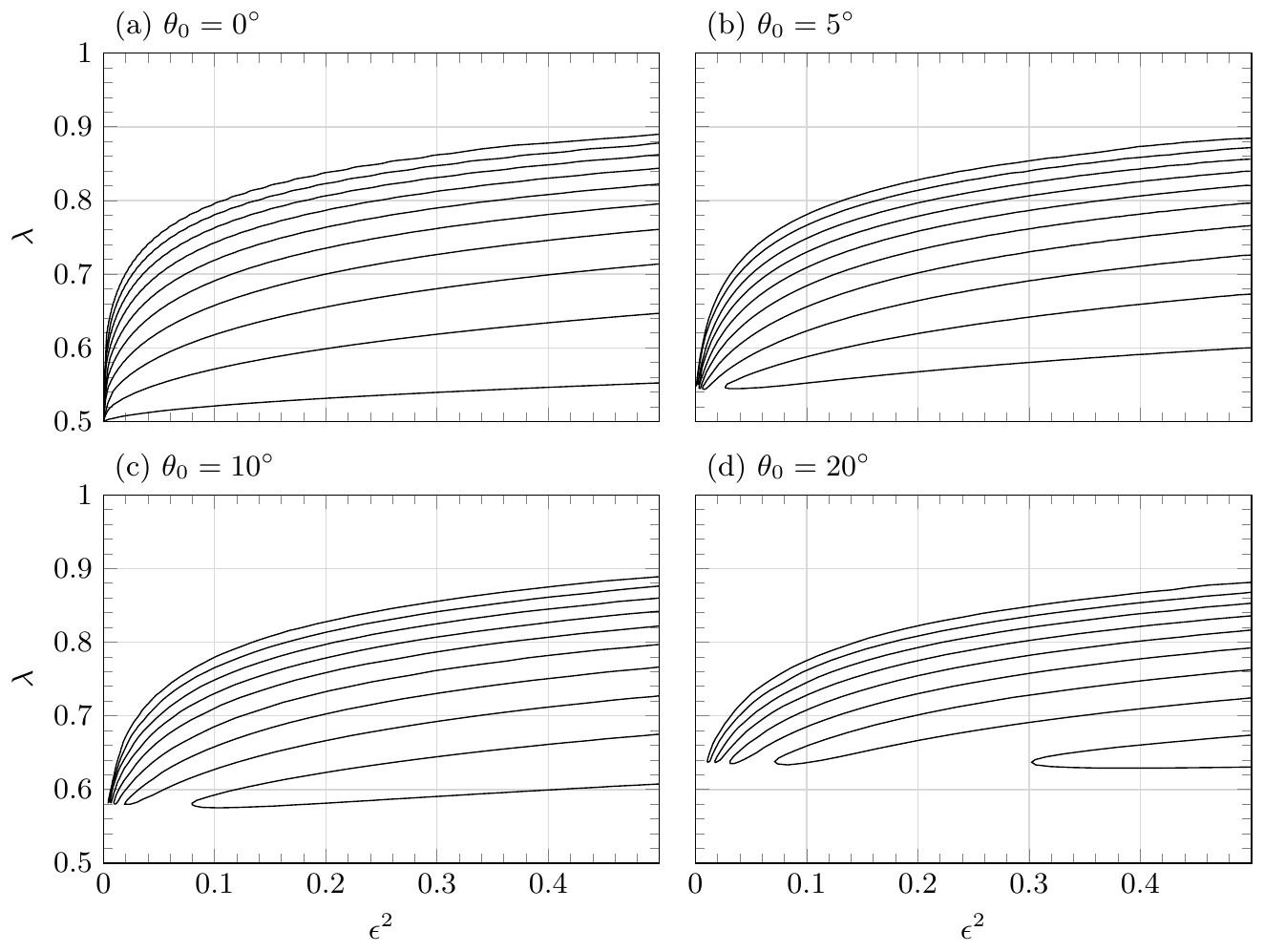}
    \caption{Bifurcation diagrams produced using (\ref{eq:selectioncondition}). The first ten selected values of $\lambda$ as functions of $\epsilon$ are shown. Subfigure (a) shows the bifurcation diagram for the channel geometry. The other subfigures (b)-(d) show the bifurcation diagram for different values of the wedge angle $\theta_0$.}
    \label{fig:BifurcationDiagram}
\end{figure}

\noindent \myblue{For each $\epsilon$ value we now use a numerical root finding scheme to find the corresponding family of $\lambda$ values that satisfy the selection condition \eqref{eq:selectioncondition}. The dependence on $\lambda$ in the selection condition \eqref{eq:selectioncondition} appears through the constants $\Lambda_1$ and $\Lambda_{\text{C}}$ as shown in Appendix \ref{app:lambda} and also through $\chi_1(0)$ and $\chi_{\text{C}}(0)$ as given in \eqref{eq:singulantintegrated}.} The selected solution families, for different values of $\theta_0$, are shown in \cref{fig:BifurcationDiagram}(a)-(d).

The figure shows that the bifurcation diagram differs qualitatively between the channel geometry with $\theta_0 = 0^\circ$ and the general wedge geometry with $\theta_0>0$. For the channel geometry each branch, $\lambda_n(\epsilon)$, continues to exist as $\epsilon \rightarrow 0$. However, for the wedge geometry we see that the branches merge and disappear in pairs and no solutions will exist when $\epsilon =0$. The lowest two branches, $\lambda_1(\epsilon)$ and $\lambda_2(\epsilon)$, merge at the greatest value of $\epsilon$ and then merging continues to happen between higher branches as $\epsilon \rightarrow 0.$ Each time such a merge happens, the solutions corresponding to those two branches will cease to exist. In practice, \myblue{we hypothesise that} the existence of the self-similar solution is lost due to a tip splitting instability which occurs when the flow rate or surface tension reaches a critical value. In \cref{fig:BifurcationDiagram} we see that for smaller wedge angles the tip splitting will occur at a smaller value of the effective surface tension $\epsilon$. In the limit $\theta_0 \rightarrow 0$, the surface tension value at which the branch merging occurs approaches zero.

As noted in \emph{e.g.} \cite{amar1991}, it is expected that, as is the case in the channel geometry, only the $\lambda_1$ branch is stable. Then the merging of the $\lambda_1$ and $\lambda_2$ branches as $\ep \to 0$ is associated with a loss of stable solutions, resulting in a tip splitting instability.

\begin{figure}
    \centering
    %\begin{preview}
    %\input{TIKZ/LambdaLimitPlots.tex}
    %\end{preview}
    \includegraphics{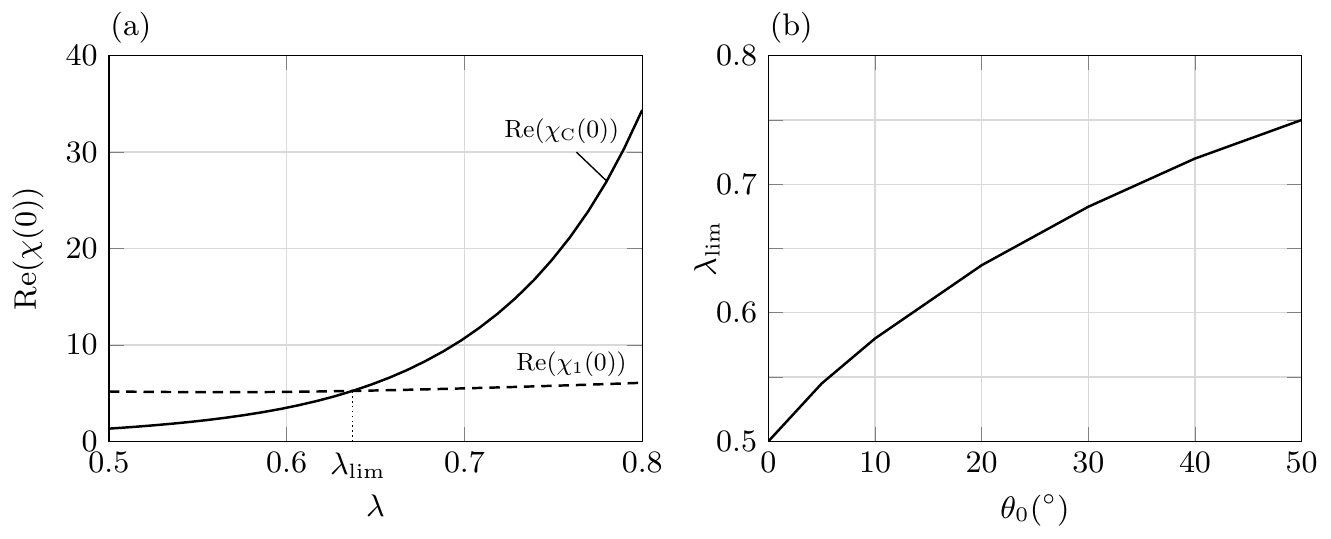}
    \caption{(a) Relative sizes of the real parts of $\chi_1(0)$ and $\chi_{\text{C}}(0)$ for different values of $\lambda$ in the $\theta_0 = 20^\circ$ case. The $\lambda$ value where $\mathrm{Re}(\chi_{\text{C}}(0)) =\mathrm{Re}(\chi_1(0))$ is denoted $\lambda_{\text{lim}}$. Above this value the selection condition \eqref{eq:selectioncondition} can be satisfied. (b) The limiting $\lambda$ value, $\lambda_{\text{lim}}$, in the small surface tension limit as a function of the wedge angle.}
    \label{fig:Realchi}
\end{figure}

The key observation is that the relative size of the terms in the selection condition \eqref{eq:selectioncondition}, provides an approximate criterion for the existence of solutions. We recall that as $\ep \to 0$, the classic Saffman-Taylor fingering in a channel involves $\lambda = \lambda_i \to \lambda_{\text{lim}} = 0.5$ for all indices $i$. It can be asked whether a similar limiting value of $\lambda$ is reached as $\ep \to 0$ and $i \to \infty$ for general $\theta_0$. 

In \cref{fig:Realchi} we plot the real parts of $\chi_1(0)$ and $\chi_{\text{C}}(0)$ for the $\theta_0 = 20^\circ$ case. As indicated in the figure, for $\lambda > \lambda_{\text{lim}}$, then $0 < \Re\chi_1(0) < \Re\chi_\text{C}(0)$. Therefore, from the selection condition, the contribution from the central Stokes line is exponentially dominated by the contribution from the `1' Stokes line(s). 

When $\mathrm{Re}(\chi)$ is smaller then the corresponding $\e^{-\mathrm{Re}(\chi)/\epsilon}$ term will dominate in the selection condition and so \cref{fig:Realchi} shows that for smaller $\lambda$ values the contribution from the `C' Stokes line dominates and for larger $\lambda$ values the contribution from the `1' Stokes line dominates, in the small $\epsilon$ limit. When the contribution from the `1' Stokes line dominates then the exponentially small oscillations can be switched off by the symmetric `2' Stokes line and solutions will exist, similarly to the channel case. However, when the `C' Stokes line dominates then in the small $\epsilon$ limit it becomes impossible to satisfy the selection condition and no solutions will exist. The $\lambda$ value at which $\mathrm{Re}(\chi_1(0)) = \mathrm{Re}(\chi_{\text{C}}(0))$ gives the limiting $\lambda$ value, $\lambda_{\text{lim}}$. The $\lambda_{\text{lim}}$ values as a function of $\theta_0$ are shown in \cref{fig:Realchi}.

\section{Conclusion}

\noindent We have used exponential asymptotics to derive the selection mechanism in the small surface tension limit for the divergent Saffman-Taylor fingers in a wedge of an arbitrary angle. The selection mechanism is based on the requirement for the exponentially small contributions, which are switched on at Stokes lines, to satisfy the farfield boundary conditions. This Stokes line structure in the complex plane relies on an understanding of the singularities in the analytic continuation of the leading-order solution. Here we must obtain this using a numerical scheme for the analytic continuation. We find the countable family of $\lambda(\epsilon)$ values and associated selected solutions that satisfy the selection condition. The bifurcation diagrams are plotted in \cref{fig:BifurcationDiagram} and show the merging of branches of $\lambda$ values as the surface tension parameter, $\epsilon$, is decreased. \myblue{We hypothesise that} the loss of existence of solutions through this branch merging relates to the tip splitting instability observed in the time-dependent flows in a circular geometry.

\section{Discussion} \label{sec:discuss}

\noindent There are a number of interesting extensions to the work in this paper in the fields of exponential asymptotics and Hele-Shaw problems.

In order to verify the correctedness of the asymptotic predictions, we have used some of the early numerical work of \cite{amar1991}. As it turns out, there are a few aspects of numerical computations of the governing equations \eqref{eq:bernoulliequation}, \eqref{eq:hilberttransformequation} and \eqref{eq:xyequation} that render it more challenging. In a forthcoming work, we will present a specialised scheme that solves for the free-surface profile of the fingers.

One of the main difficulties with the analysis in this paper arises from the form of the leading-order solutions. Traditionally in problems considered with exponential asymptotics, the leading-order solution will have a simple closed form; then, locations and strengths of the singularities in the complex plane are easily identified. For this problem, the leading-order solution is written in terms of special functions [cf. \eqref{eq:xyhypergeometric} and \eqref{eq:qthhypergeometric}]; however the singularities of these expressions are not so easily obtained. In practice, we use a numerical scheme to analytically continue the leading-order solutions into the complex $\zeta$-plane and search for singularities. This numerical scheme could equally be used to analytically continue a fully numerical leading-order solution into the complex plane. Similar numerical analytic continuation schemes can be found in \cite{Chandler2018, Crew2016}. With the advancement of such schemes, in the field of exponential asymptotics we are no longer restricted only to problems with a simple analytic leading-order solution.

To complete the understanding of the tip splitting behaviour in the circular geometry it will be necessary to check the stability of the branches plotted in the bifurcation diagram (\cref{fig:BifurcationDiagram}). For the Saffman-Taylor problem in the channel geometry, \cite{Tanveer1987} showed that only the lowest branch, $\lambda_1$, is linearly stable. We conjecture that the same will be true for the wedge geometry and so the merging of branches $\lambda_1$ and $\lambda_2$ will correspond to the loss of existence of any solutions which could be observed physically. \myblue{We have written time-dependent code for the circular geometry based on the numerical method described in \cite{Dallaston2014} and we hope that this will provide an insight into the stability of the branches observed in the bifurcation diagram for the wedge problem.}

Finally, there are many recent experimental and numerical results for different Hele-Shaw problems, some of which appear to have similar selected branches of permitted solutions \citep{Keeler2022, Gaillard2021}. We expect that the techniques used here can also be applied to these problems to derive selection conditions, governed by exponentially small components of the solutions. In particular, the ability to do exponential asymptotics without a simple analytic leading-order solution will be necessary to make progress in these problems.

\mbox{}\par
\noindent \textbf{Acknowledgements.} We especially thank Jon Chapman (Oxford) for helping us navigate the formulation of this problem and getting us started on the analysis.  His time and effort is very much appreciated.  Further, we thank Josh Shelton (Bath) and John King (Nottingham) for many stimulating and helpful discussions. We would like to thank the Isaac Newton Institute for Mathematical Sciences, Cambridge, for support and hospitality during the programmes ``Complex Analysis: Techniques, Applications and Computations'' and ``Applicable Resurgent Asymptotics: Towards a Universal Theory'', where some work on this paper was undertaken (EPSRC grant no. [EP/R014604/1]). Some work on this paper was also conducted while visiting the Okinawa Institute of Science and Technology (OIST) through the Theoretical Sciences Visiting Program (TSVP). CA and PHT gratefully acknowledge support by the Engineering and Physical Sciences Research Council (EPSRC) [EP/V012479/1]. CJL gratefully acknowledges support by Australian Research Council Discovery Project DP190101190. For the purpose of open access, the authors have applied a Creative Commons Attribution (CC-BY) licence  to any Author Accepted Manuscript version arising.

\mbox{}\par

\noindent \textbf{Declaration of Interests.} The authors report no conflict of interest.
% 

% \bibliographystyle{jfm}
% \nocite{*}
% \bibliography{mybib}

\appendix

\section{Further details on governing equations} \label{app:formulation}

\myblue{\noindent In this section, we briefly present the details for deriving the governing equations. The fluid within the Hele-Shaw cell is governed by Stokes flow, and we consider the setup as shown in \cref{fig:HeleShaw}(a). We shall use bars for dimensional quantities, including Cartesian coordinates $(\bar{x},\bar{y})$ and polar coordinates $(\bar{r},\theta)$. Let ${\bf \bar{q}}$ be the dimensional velocity averaged over the gap. The fluid pressure, $\bar{p}$, is given by Darcy flow, ${\bf \bar{q}}=-(b^2/12\mu)\bar\nabla \bar p$ where $b$ is the Hele-Shaw cell gap thickness and $\mu$ is the viscosity. We introduce a potential via $\bar{\phi}=-(b^2/12\mu)p$. Then following the standard derivation via Stokes flow \citep{ockendon1995viscous}, we see the viscous fluid within the Hele-Shaw cell is governed by Laplace's equation, $\nabla^2\bar\phi=0$, for a velocity potential with ${\bf \bar{q}}=\bar\nabla \bar\phi$.} 
    
% On the interface between the viscous and inviscid fluids we must impose a kinetic boundary condition, which states that the interface moves with same velocity as the fluid, and the dynamic boundary condition, which relates fluid pressure to the interface curvature and surface tension.  We refer to this expression as the Bernoulli equation.  Finally, for our geometry there is a boundary condition in the far field that describes how fast inviscid fluid is injected at the wedge corner (or, alternatively, how fast viscous fluid is extracted at infinity).  All of these conditions are written out in Appendix~\ref{app:formulation} in dimensional and dimensionless form. }

\myblue{Inviscid fluid is injected at the origin and the interface between viscous and inviscid fluids is given by $\bar{r}=\bar{R}(\theta,\bar{t})$. The full set of governing equations, including interfacial and flux conditions are
\begin{subequations}
\begin{align}
\bar{\nabla}^2\bar{\phi}&=0 && \text{in the fluid}, \\
\bar{v}_n&=\frac{\partial\bar{\phi}}{\partial \bar{n}} && \text{on $\bar{r}=\bar{R}(\theta,\bar{t})$}, \label{eq:kin_tmp1} \\
\bar{\phi} - \bar{\phi_0}&=-\frac{b^2\sigma}{12\mu}\,\bar{\kappa} && \text{on $\bar{r}=\bar{R}(\theta,\bar{t})$}, \label{eq:dyn_tmp1} \\
\frac{\partial\bar{\phi}}{\partial\bar{n}} &=0 && \text{on $\bar{y}=\pm \bar{x}\tan (\theta_0/2)$,}  \label{eq:wall_tmp1} \\
\frac{\partial\bar{\phi}}{\partial\bar{r}} &\sim \frac{\bar{Q}}{2\pi b\bar{r}} && \text{as $\bar{r}\rightarrow\infty$} \label{eq:eq:flux_tmp1},
\end{align}
\end{subequations}
where $\sigma$ is the dimensional surface tension, $\bar{v}_n$ is the velocity normal to the interface, $\bar{\kappa}$ is the curvature of the interface, and $\bar\phi_0$ an arbitrary shift of the potential. Equations \eqref{eq:kin_tmp1} and \eqref{eq:wall_tmp1} correspond to kinematic conditions on the interface and wall, respectively, \eqref{eq:dyn_tmp1} to the dynamic boundary condition on the interface, and \eqref{eq:eq:flux_tmp1} the fluid injection condition for a dimensional source strength $\bar{Q}$.}

\subsection*{Temporal scaling and non-dimensionalisation}

We scale time by a yet-to-be-specified time-scale, $T$, so that dimensionless time is $t=\bar{t}/T$ and introduce a length scale, $R_0$, which is the initial distance ($t = 0$) between the tip of the bubble and the apex. We shall apply a time-dependent stretching of the domain via
\begin{equation}
\bar{R}(\theta,\bar{t})=\mathrm{A}(t)\bar{R}(\theta,0) = \mathrm{A}(t)R_0\hat{R}(\theta),
\end{equation}
where $A(t)$ is to be specified. Our dimensionless variables are now denoted without the overbar and we we write: 
\begin{equation}
(\hat{x},\hat{y})=\frac{(\bar{x},\bar{y})}{R_0\mathrm{A}(t)}, \quad \hat{r}=\frac{\bar{r}}{R_0\mathrm{A}(t)}, \qquad \phi= \frac{T\bar{\phi}}{R_0^2}, 
\end{equation}
and so forth. The governing equations are now
\begin{subequations} \label{eq:tochoose}
\begin{align}
\hat{\nabla}^2\phi &=0 && \text{in the fluid}, \\
\hat{\nabla}\phi\cdot {\bf \hat n} &= \mathrm{A}(t)\mathrm{A}'(t){\bf \hat R}\cdot{\bf \hat n} && \text{on $\hat{r} = \hat{R}(\theta)$}, \\
\phi - \phi_0&=-\frac{b^2T\sigma}{12\mu R_0^3\mathrm{A}(t)}\,\hat\kappa && \text{on $\hat{r} = \hat{R}(\theta)$}, \\
\frac{\partial{\phi}}{\partial \hat{n}} &=0 && \text{on $\hat{y}=\pm \hat{x}\tan (\theta_0/2)$,} \\
\frac{\partial\phi}{\partial \hat{r}} &\sim \frac{\bar{Q}\,T}{2\pi b R_0^2 \hat{r}} &&\text{as $\hat{r}\rightarrow\infty$.}
\end{align}
\end{subequations}

We review two ways to further reduce the above set of equations via the choice of $\timeA$. The first is to define $\hat{\phi}=\mathrm{A}(t)(\phi-\phi_0)$ and
\begin{equation}
\mathrm{A}(t)^2\mathrm{A}'(t)=1, \qquad \hat{\sigma}=\frac{b^2T\sigma}{12\mu R_0^3}, \qquad \frac{\bar{Q}T\mathrm{A}(t)}{b R_0^2}=1.
\end{equation}
Then the governing equations \eqref{eq:tochoose} become
\begin{subequations} \label{nearfinal_nondim}
\begin{align}
\hat{\nabla}^2\hat{\phi} &=0  && \text{in the fluid}, \label{eq:laplacephi1} \\
\hat{\nabla}\hat{\phi}\cdot {\bf \hat n} &= {\bf \hat R}\cdot{\bf \hat n} && \text{on the interface},\label{eq:kineticphi1} \\
\hat\phi &=-\hat{\sigma}\hat{\kappa} && \text{on the interface}, \label{eq:dynamicphi1} \\
\frac{\partial \hat{\phi}}{\partial \hat{n}} &=0 && \text{on $\hat{y}=\pm \hat{x}\tan(\theta_0/2)$,} \label{eq:wallphhi1} \\
\frac{\partial \hat{\phi}}{\partial \hat{r}} &\sim \frac{1}{2\pi \hat{r}} && \text{as $\hat{r}\rightarrow\infty$}. \label{eq:farfieldphi1}
\end{align}
\label{eq:governingequationsphi1}
\end{subequations}
\noindent In this way, our domain stretching function and time-scale become
\begin{equation}
\mathrm{A}(t)=(3t+1)^{1/3} \quad \text{and} \quad T=\frac{b R_0^2}{\bar{Q}(0)},
\end{equation}
our surface tension parameter $\hat{\sigma}$ is a constant, and the flow-rate must be
\begin{equation}
\bar{Q}(t)=\frac{\bar{Q}(0)}{(3t+1)^{1/3}}.
\end{equation}
This way of reducing the equations has the attraction of keeping the surface tension constant (which is what happens in reality), while also requiring a non-constant flow rate with a $t^{1/3}$ scaling.  Such a flow-rate is perfectly realisable in practice \citep{Li2009}.

The other way of reducing the equations is to set $\hat{\phi}=\phi-\phi_0$ and define 
\begin{equation} \label{eq:benamar_paramchoice}
\mathrm{A}(t)\mathrm{A}'(t)=1, \qquad \hat{\sigma}=\frac{b^2T\sigma}{12\mu R_0^3\mathrm{A}(t)}, \qquad \frac{\bar{Q}T}{b R_0^2}=1,
\end{equation}
so the governing equations \eqref{eq:tochoose} are the same as above \eqref{eq:governingequationsphi1}.  This time we have
\begin{equation}
\mathrm{A}(t)=(2t+1)^{1/2} \qquad \text{and} \quad T=\frac{b R_0^2}{\bar{Q}}.
\end{equation}
This approach has the advantage of keeping the flow-rate, $\bar{Q}$, constant; this is the most common realisation in an experimental set-up, but has the consequence of requiring the consideration of a time-dependent surface tension parameter $\hat{\sigma}$---conceptually, we may consider $\sigma$ as slowly varying, and even treated as  constant within the context of a simulation (see further discussion in \citealt{amar1991}).

\section{Deriving the power $\gamma$} \label{app:gamma}

\noindent In this section we derive the constants $\gamma_1$ and $\gamma_{\text{C}}$, which arise in \eqref{eq:selectioncondition}, by matching the strengths of the singularities in the inner region. Firstly, we introduce a new variable, $Y = \zeta - \zeta_*$, where $\zeta_*$ is one of the singularities, $\zeta_1$ or $\zeta_{\text{C}}$.

It can be verified, either through dominant balance, or via the  numerical continuation of \cref{sec:analyticcontinuationandsingularities} that all the complex plane singularities of the leading-order speed, $q_0$, are square root singularities. Hence, $q_0 \sim Y^{-\frac{1}{2}}$ as $Y \rightarrow 0$. Close to the singularity, the Hilbert transform in the boundary integral equation will be subdominant and so we can use $q_0 - i\tau_0 = \mathcal{H}[\tau_0]$ and the behaviour of $q_0$ to derive the local $\tau_0$ behaviour,
\begin{equation}
\tau_0 \sim \frac{\i}{2}\log Y \quad \text{as} \quad Y \rightarrow 0,
\end{equation}
and hence $\mathrm{e}^{\i\tau_0} \sim Y^{-1/2}$ in this limit. The asymptotic singular behaviour for other key variables is given as $Y \to 0$ by,
\begin{subequations}
\begin{align}
x_0 + \i &y_0 \sim \int_Y^{-\zeta_*} 1 \: \de{\tilde{Y}} = \mathcal{O}(Y), \qquad P = \mathcal{O}(1), \\
&I(\zeta) = \mathcal{O}(1), \qquad Q = \mathcal{O}\left(Y^{-3/8}\right).
\end{align}
\end{subequations}
The local expression for the singulant equation \eqref{eq:singulant} is given by,
\begin{equation}
    \chi' = \frac{2\sqrt{\i}\e^{-\i\tau_0/2}}{q_0P}\e^{{3\theta_0}x_0/4} = \mathcal{O}\left(Y^{3/4}\right) \quad \mathrm{as} \quad Y \to 0.
\end{equation}
Integrating this expression gives $\chi = \mathcal{O}(Y^{7/4})$ as $Y \to 0$. The factorial-over-power ansatz \eqref{eq:factorialoverpower} then gives the strength of the singularity in the later order terms,
\begin{equation}
q_n \sim \frac{Q\Gamma(2n+\gamma)}{\chi^{2n+\gamma}} = \mathcal{O}\left(Y^{-3/8}Y^{-(2n+\gamma)7/4}\right) \quad \text{as} \quad Y\rightarrow 0.
\label{eq:latetermssingularbehaviour}
\end{equation}
Now we take $n=0$ and choose $\gamma$ so that the correct predicted singularity strength is given for $q_0.$ Although the factorial-over-power ansatz was posed for the limit $n \rightarrow \infty,$ the singularity strength at lower orders still needs to match for the two expressions to be consistent. The $\gamma$ that satisfies this condition is,
\begin{equation}
\gamma = \frac{1}{14}.
\end{equation}
This analysis is valid for each singularity, so $\gamma_1 = \gamma_{\text{C}} = 1/14$, irrespective of the wedge angle $\theta_0$.

\section{Deriving the pre-factor $\Lambda$ } \label{app:lambda}

\noindent To find $\Lambda$, which appears in the late-order expression \eqref{eq:lateorderq} and also the exponential switching \eqref{eq:selectioncondition} we consider the solution in an inner region near the singularity at $\zeta = \zeta^*$. We define a new inner variable $\nu$ such that $\zeta - \zeta_* = \epsilon^\alpha \nu$. The value of $\alpha$ determines the width of the inner region. From (\ref{eq:latetermssingularbehaviour}) with $\gamma = 1/14$, the local behaviour of the late terms near the singularity is given by
\begin{equation}
    q_n = \mathcal{O}\left(\left(\epsilon^\alpha \nu\right)^{-3/8 - (2n+\frac{1}{14})7/4}\right) \quad \text{as} \quad \epsilon \rightarrow 0.
\end{equation}
We locate the inner region by identifying where terms in the power series (\ref{eq:qtaupowerseries}) reorder, and the power series expansion is therefore no longer valid. This occurs when $\epsilon^{2n}q_n$ is approximately $\mathcal{O}(q_0)$ as $n \rightarrow \infty$. That is,
\begin{equation}
    2n + \alpha\left(-\frac{3}{8} - \frac{7n}{2} - \frac{1}{8}\right) =  -\frac{\alpha}{2},
\end{equation} which gives $\alpha = {4}/{7}$. The correct scaling for the inner region is thus
\begin{equation}
    \zeta = \zeta_* + \epsilon^{4/7} \nu,
\end{equation}
which is identical to the scaling for the channel geometry \citep{chapman1999}.

We also require the local behaviour of the dependent variables near the singularities. This is done numerically using the analytic continuation scheme of \cref{sec:analyticcontinuationandsingularities}. Beginning from the free surface, we solve for the analytic continuation along a path that approaches the singularity, $\zeta_*$. Once done, the local behaviour of the dependent variables can be determined by numerical fitting to log-log plots. In particular, we find: 
\begin{subequations}
\begin{align}
&x_0 \sim b_{1*} + b_{2*} \epsilon^{2/7}\nu^{1/2} + \mathcal{O}\left(\epsilon^{4/7}\right), \\
&q_0 \sim c_{1*}\epsilon^{-2/7}\nu^{-1/2} + c_{2*}\epsilon^{2/7}\nu^{1/2} + \mathcal{O}\left(\epsilon^{6/7}\right), \\
&\e^{i\tau_0} \sim d_{1*}\epsilon^{-2/7}\nu^{-1/2} + d_{2*} \epsilon^{2/7}\nu^{1/2} + \mathcal{O}\left(\epsilon^{6/7}\right).
\end{align}
\end{subequations}
Here the coefficients $b_{1*}, b_{2*}, c_{1*}, c_{2*}, d_{1*}, d_{2*}$ are constant with respect to the independent variables, but they depend on the parameters $\theta_0$ and $\lambda$ and also vary between the different singularities. We indicate the choice of singularity using the $*$ subscript. The constants are typically complex. Selected values of the constants as $\theta_0$ and $\lambda$ vary are tabulated in \cref{fig:ConstantsLeft} and \cref{table:zetaCinnerconstants} for the `1' and `$\text{C}$' singularities respectively.

\begin{landscape}
\pagestyle{lscape}
\begin{figure}
    \centering
    %\begin{preview}
    \begin{tikzpicture}
\tikzstyle{circstyle}=[circle, draw=black, fill=white, inner sep=0pt, minimum size=0.2cm, line width=0.8pt]
\tikzstyle{secout}=[circle, draw=black, fill=black, inner sep=0pt, minimum size=0.22cm, line width=0.8pt]
\tikzstyle{secin}=[circle, draw=white, inner sep=0pt, minimum size=0.15cm, line width=0.8pt]

\begin{axis}[
    width=20cm,
    height=12.5cm,
    xmin={0},
    xmax={1.75},
    ymin={0},
    ymax={3.5},
    xlabel={$\mathrm{Re}(\zeta)$},
    ylabel={$\mathrm{Im}(\zeta)$},
    xtick = { 0, 0.2,0.4,0.6, 0.8,1, 1.2,1.4, 1.6},
    ytick = { 0, 0.5, 1,1.5,  2,2.5, 3}
]

\addplot[mark=none, gray!70] coordinates {(0,1.2222) (0.2699,1.09631)  (0.5801,1.0323) (0.9008,0.9630) (1.2451,0.8552)};
\addplot[mark=none, gray!70] coordinates { (0,1.5)  (0.2533, 1.3976) (0.5854, 1.3237)  (0.9464, 1.2697) (1.3400, 1.1945)};
\addplot[mark=none, gray!70] coordinates {(0, 1.8571)  (0.2580, 1.7891)  (0.5875, 1.7251) (0.9748, 1.6821) (1.4101, 1.6397)};
\addplot[mark=none, gray!70] coordinates {(0, 2.3333)(0.2702, 2.2989)  (0.6006, 2.2656) (0.9963, 2.2453)  (1.4566, 2.2374)};
\addplot[mark=none, gray!70] coordinates { (0, 3) (0.2845,3.0007)  (0.6211, 3.0068)  (1.0191, 3.0254) (1.4866, 3.0617)};

\addplot[mark=none, gray!70] coordinates {(0,1.2222) (0,1.5) (0, 1.8571) (0, 2.3333) (0, 3)};
\addplot[mark=none, gray!70] coordinates {(0.2699,1.09631) (0.2533, 1.3976) (0.2580, 1.7891) (0.2702, 2.2989) (0.2845,3.0007)};
\addplot[mark=none, gray!70] coordinates {(0.5801,1.0323) (0.5854, 1.3237) (0.5875, 1.7251) (0.6006, 2.2656) (0.6211, 3.0068)};
\addplot[mark=none, gray!70] coordinates {(0.9008,0.9630) (0.9464, 1.2697) (0.9748, 1.6821) (0.9963, 2.2453)  (1.0191, 3.0254)};
\addplot[mark=none, gray!70] coordinates {(1.2451,0.8552) (1.3400, 1.1945) (1.4101, 1.6397) (1.4566, 2.2374)  (1.4866, 3.0617)};

\node [anchor=south west, black, scale=1] at (axis cs:-0.005,3.03) {$0^\circ$};
\node [anchor=south west, black, scale=1] at (axis cs:0.25,3.03) {$10^\circ$};
\node [anchor=south west, black, scale=1] at (axis cs:0.6,3.04) {$20^\circ$};
\node [anchor=south west, black, scale=1] at (axis cs:0.98,3.06) {$30^\circ$};
\node [anchor=south east, black, scale=1] at (axis cs:1.52,3.1) {$40^\circ$};

\node [anchor=south west, black, scale=1] at (axis cs:1.257,0.82) {$0.55$};
\node [anchor= west, black, scale=1] at (axis cs:1.345,1.22) {$0.60$};
\node [anchor= west, black, scale=1] at (axis cs:1.41,1.68) {$0.65$};
\node [anchor= west, black, scale=1] at (axis cs:1.46,2.28) {$0.70$};
\node [anchor= west, black, scale=1] at (axis cs:1.492,3.1) {$0.75$};

\node [align=left,anchor= north west, gray, scale=0.8] at (axis cs:0,1.2222) {(1,\\ 0.55+0.55i,\\ 1.11+1.11i)};
\node [align=left,anchor= north west, gray, scale=0.8] at (axis cs:0.2699,1.09631) {(1-0.03i, \\0.58+0.49i, \\1.06+1.16i)};
\node [align=left,anchor= north west, gray, scale=0.8] at (axis cs:0.5801,1.0323) {(0.99-0.04i, \\0.67+0.37i, \\1.12+1.04i)};
\node [align=left,anchor= north west, gray, scale=0.8] at (axis cs:0.9008,0.9630) {(0.97-0.05i, \\0.79+0.27i, \\1.21+0.94i)};
\node [align=left,anchor= north west, gray, scale=0.8] at (axis cs:1.2451,0.8552) {(0.93-0.05i, \\0.92+0.17i, \\1.34+0.83i)};
\node [align=left,anchor= north west, gray, scale=0.8] at (axis cs:0,1.5) {(1,\\ 0.61+0.61i,\\ 1.22+1.22i)};
\node [align=left,anchor= north west, gray, scale=0.8] at (axis cs:0.26, 1.3976) {(1-0.03i, \\0.68+0.56i, \\1.29+1.28i)};
\node [align=left,anchor= north west, gray, scale=0.8] at (axis cs:0.5854, 1.3237) {(0.98-0.05i, \\0.74+0.47i, \\1.28+1.23i)};
\node [align=left,anchor= north west, gray, scale=0.8] at (axis cs:0.9464, 1.2697) {(0.96-0.06i, \\0.85+0.37i, \\1.35+1.13i)};
\node [align=left,anchor= north west, gray, scale=0.8] at (axis cs:1.3400, 1.1945) {(0.92-0.06i, \\0.98+0.27i, \\1.47+1.03i)};
\node [align=left,anchor= north west, gray, scale=0.8] at (axis cs:0, 1.8571) {(1,\\ 0.68+0.68i,\\ 1.36+1.36i)};
\node [align=left,anchor= north west, gray, scale=0.8] at (axis cs:0.2580, 1.7891) {(0.99-0.03i, \\0.76+0.63i, \\1.45+1.39i)};
\node [align=left,anchor= north west, gray, scale=0.8] at (axis cs:0.5875, 1.7251) {(0.97-0.05i, \\0.83+0.56i, \\1.49+1.39i)};
\node [align=left,anchor= north west, gray, scale=0.8] at (axis cs:0.9748, 1.6821) {(0.95-0.06i, \\0.93+0.48i, \\1.53+1.34i)};
\node [align=left,anchor= north west, gray, scale=0.8] at (axis cs:1.4101, 1.6397) {(0.91-0.07i, \\1.05+0.39i, \\1.64+1.25i)};
\node [align=left,anchor= north west, gray, scale=0.8] at (axis cs:0, 2.3333) {(1,\\ 0.76+0.76i,\\ 1.53+1.53i)};
\node [align=left,anchor= north west, gray, scale=0.8] at (axis cs:0.2702, 2.2989) {(0.98-0.03i, \\0.84+0.71i, \\1.61+1.54i)};
\node [align=left,anchor= north west, gray, scale=0.8] at (axis cs:0.6006, 2.2656) {(0.96-0.05i, \\0.92+0.66i, \\1.68+1.55i)};
\node [align=left,anchor= north west, gray, scale=0.8] at (axis cs:0.9963, 2.2453) {(0.94-0.06i, \\1.02+0.59i, \\1.75+1.53i)};
\node [align=left,anchor= north west, gray, scale=0.8] at (axis cs:1.4566, 2.2374) {(0.90-0.07i, \\1.13+0.52i, \\1.84+1.48i)};
\node [align=left,anchor= north west, gray, scale=0.8] at (axis cs:0, 3) {(1,\\ 0.87+0.87i,\\ 1.73+1.73i)};
\node [align=left,anchor= north west, gray, scale=0.8] at (axis cs:0.2845,3.0007) {(0.98-0.02i, \\0.94+0.82i, \\1.82+1.74i)};
\node [align=left,anchor= north west, gray, scale=0.8] at (axis cs:0.6211, 3.0068) {(0.96-0.04i, \\1.03+0.77i, \\1.90+1.75i)};
\node [align=left,anchor= north west, gray, scale=0.8] at (axis cs:1.0191, 3.0254) {(0.93-0.06i, \\1.12+0.72i, \\1.98+1.75i)};
\node [align=left,anchor= north west, gray, scale=0.8] at (axis cs:1.4866, 3.0617) {(0.89-0.07i, \\1.23+0.66i,\\ 2.07+1.74i)};

\draw [-to, black](0.75,3.29) -- (0.95,3.32);
\node [anchor = south, black, rotate=2, scale=1] at (axis cs: 0.65, 3.2) {$\theta_0$ increasing};
\draw [-to,black](1.644,1.6) -- (1.665,2.1);
\node [anchor = south, black, rotate=80, scale=1] at (axis cs: 1.65, 1.3) {$\lambda$ increasing};

\addplot+[mark=*, color=black, mark options={fill=white}, inner sep=0pt, minimum size=0.1cm, mark size=2.5pt, line width=1pt, only marks] 
  coordinates
  {(0.2699,1.09631)  (0.5801,1.0323)  (0.9008,0.9630) (1.2451,0.8552)   (0.2533, 1.3976) (0.5854, 1.3237)  (0.9464, 1.2697)  (1.3400, 1.1945)  (0.2580, 1.7891)  (0.5875, 1.7251)  (0.9748, 1.6821)  (1.4101, 1.6397)   (0.2702, 2.2989) (0.6006, 2.2656)  (0.9963, 2.2453)  (1.4566, 2.2374)   (0.2845,3.0007) (0.6211, 3.0068) (1.0191, 3.0254) (1.4866, 3.0617) };

\end{axis}

\node[secout, scale=0.5] at (0, 3.1) {};

\draw[line width=0.8pt, decorate,decoration={snake,amplitude=.7mm,segment length=3mm,post length=0mm}] (0,3.1) -- (0,11);

\node[secout, scale=0.83] at (0,3.815){};
\node[secin, scale=0.7] at (0,3.815){};
\node[secin, scale=0.5] at (0,3.815){};
\node[secin, scale=0.2] at (0,3.815){};
\node[secout, scale=0.83] at (0,4.68){};
\node[secin, scale=0.7] at (0,4.68){};
\node[secin, scale=0.5] at (0,4.68){};
\node[secin, scale=0.2] at (0,4.68){};
\node[secout, scale=0.83] at (0,5.79){};
\node[secin, scale=0.7] at (0,5.79){};
\node[secin, scale=0.5] at (0,5.79){};
\node[secin, scale=0.2] at (0,5.79){};
\node[secout, scale=0.83] at (0,7.28){};
\node[secin, scale=0.7] at (0,7.28){};
\node[secin, scale=0.5] at (0,7.28){};
\node[secin, scale=0.2] at (0,7.28){};
\node[secout, scale=0.83] at (0,9.37){};
\node[secin, scale=0.7] at (0,9.37){};
\node[secin, scale=0.5] at (0,9.37){};
\node[secin, scale=0.2] at (0,9.37){};

\end{tikzpicture}
    %\end{preview}
    %\includegraphics{figures/innerconstantsleftTikz.pdf}
\caption{Locations of the `1' singularity for different values of $\lambda$ and $\theta_0$. At each singularity we have included the corresponding triplet of complex constants for the leading-order inner behaviour around the `1' singularity, $(b_1, c_1, d_1).$}
\label{fig:ConstantsLeft}
\end{figure}
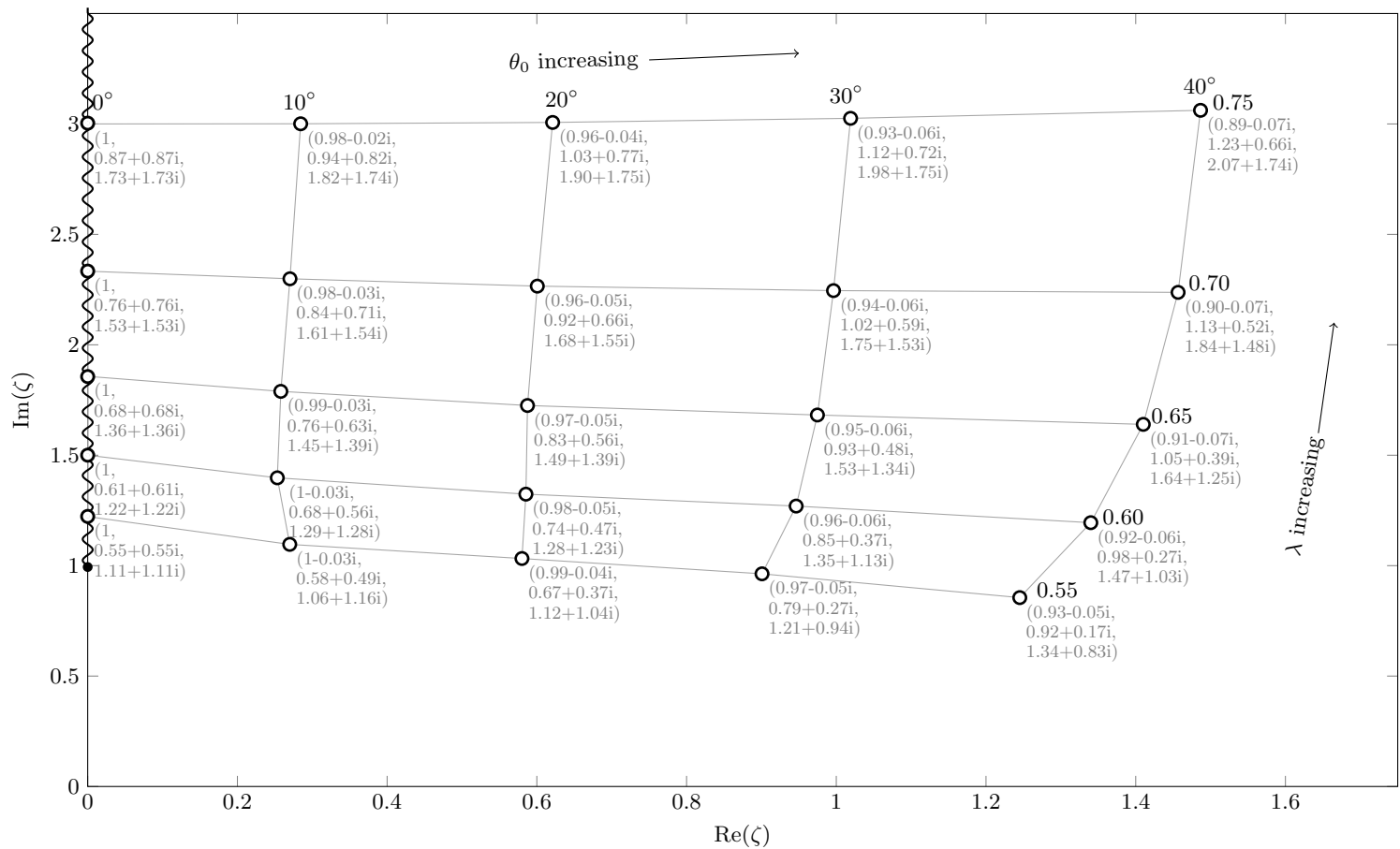
\end{landscape}

\begin{table}
\begin{center}
\begin{tabular}{ c | c c c c}
\backslashbox{$\lambda$}{$\theta_0$} & $10^{\circ}$  & $20^{\circ}$ &$30^{\circ}$  & $40^{\circ}$\\
\hline
\vspace{0.2cm}
 \shortstack[c]{0.55 \\ $\;$  \\ $\;$ \\$\;$}  &\shortstack[r]{$(1.04+0.01\i,\;$\\ $ 0.27+0.27\i,\;$\\ $0.62+0.60\i)\,$}  &\shortstack[r]{$(1.05+0.00\i,\;$\\$ 0.30+0.30\i,\;$\\$ 0.73+0.73\i)\,$} &\shortstack[r]{$(1.05+0.00\i,\;$\\$ 0.29+0.29\i,\;$\\$ 0.73+0.73\i)\,$}  &\shortstack[r]{$(1.05+0.00\i,\;$\\$ 0.26+0.27\i,\;$\\$ 0.71+0.71\i)\,$} \\ 
 \vspace{0.2cm}
 \shortstack[c]{0.60 \\ $\;$  \\ $\;$ \\$\;$} &\shortstack[r]{$(1.04+0.09\i,\;$\\$0.07+0.01\i,\;$\\$0.17+0.02\i)\, $}&\shortstack[r]{$ (1.08+0.01\i,\;$\\$ 0.21+0.21\i,\;$\\$0.54+0.53\i)\,$}&\shortstack[r]{$ (1.07+0.00\i,\;$\\$ 0.24+0.24\i,\;$\\$ 0.66+0.65\i)\,$} &\shortstack[r]{$(1.07+0.00\i,\;$\\$ 0.24+0.24\i,\;$\\$ 0.69+0.68\i)\,$}\\  
 \vspace{0.2cm}
 \shortstack[c]{0.65 \\ $\;$  \\ $\;$ \\$\;$} & \shortstack[r]{$(0.98+0.06\i,\;$\\$0.02-0.04\i,\;$\\$ 0.06-0.11\i)\,$}&\shortstack[r]{$ (1.13+0.05\i,\;$\\$ 0.09+0.08\i,\; $\\$0.26+0.22\i)\,$}&\shortstack[r]{$(1.12+0.01\i,\; $\\$0.17+0.16\i,\;$\\$ 0.49+0.48\i)\,  $}&\shortstack[r]{$(1.10+0.01\i,\;$\\$ 0.20+0.19\i,\;$\\$ 0.60+0.60\i)\,$}\\
 \vspace{0.2cm}
 \shortstack[c]{0.70 \\ $\;$  \\ $\;$ \\$\;$} & \shortstack[r]{$(0.98+0.05\i,\;$\\$ 0.01-0.03\i,\;$\\$ 0.03-0.10\i)\,$}&\shortstack[r]{$(1.04+0.20\i,\;$\\$0.03-0.00\i,\;$\\$ 0.11-0.00\i)\,$ }&\shortstack[r]{$ (1.18+0.05\i,\;$\\$ 0.09+0.08\i,\;$\\$0.29+0.26\i)\, $}&\shortstack[r]{$(1.16+0.02\i,\;$\\$ 0.13+0.13\i,\;$\\$ 0.45+0.44\i)\,$}\\
 \shortstack[c]{0.75 \\ $\;$  \\ $\;$ \\$\;$} & \shortstack[r]{$(0.98+0.05\i,\;$\\$ 0.00-0.02\i,\;$\\$0.01-0.08\i)\,$}&\shortstack[r]{$ (0.93+0.14\i,\;$\\$0.01-0.02\i,\;$\\$ 0.06-0.07\i)\,$}&\shortstack[r]{$ (1.22+0.22\i,\;$\\$ 0.03+0.02\i,\;$\\$0.14+0.08\i)\, $}&\shortstack[r]{$(1.24+0.05\i,\;$\\$ 0.07+0.07\i,\;$\\$ 0.28+0.27\i)\,$}
\end{tabular}
\end{center}
\caption{The triplet of complex constants for the leading-order inner behaviours in the inner region about the `C' singularity for different values of $\theta_0$ and $\lambda$. }
\label{table:zetaCinnerconstants}
\end{table}

\noindent From these expansions it follows that $\tau$ can be written in terms of $q$, 
\begin{equation}
    \begin{aligned}
        \log q - \i\tau &= \hat{\mathcal{H}}[\tau_0] + \mathcal{O}(\epsilon^2) = \log q_0 - \i\tau_0 + \mathcal{O}(\epsilon^2) \\
        &= \log \left(\frac{c_{1*}}{d_{1*}}\right) + \left(\frac{c_{2*}}{c_{1*}} - \frac{d_{2*}}{d_{1*}}\right)\epsilon^{4/7}\nu + \mathcal{O}\left(\epsilon^{8/7}\right),
    \end{aligned}
\end{equation}
and $r$ can be expanded,
\begin{equation}
    r(x) = r(x_0) + \mathcal{O}(\epsilon^2) = r(b_{1*}) - r(b_{1*})\frac{\theta_0}{2}\epsilon^{2/7}\nu^{1/2} + \mathcal{O}(\epsilon^{4/7}).
\end{equation}

\noindent Next we will derive the inner equation from the Bernoulli equation in terms of the inner variable for $q$, which is defined as, 
\begin{equation}
    q = c_{1*} \epsilon^{-2/7} \nu^{-1/2} q_{in}.
\end{equation}
The analytically continued Bernoulli equation \eqref{eq:bernoulliAC} is,
\begin{equation}
    \epsilon^2 \frac{P}{4}\frac{\partial}{\partial \zeta}\left(r(x)\left[Pq\frac{\partial \tau}{\partial \zeta} - \t \sin\tau\right]\right) = \frac{2}{\pi}\int_{-\infty}^0 \frac{K(\tilde{\zeta}) \tilde{\zeta}}{\zeta^2 - \tilde{\zeta}^2}\de{\tilde{\zeta}} + \i K(\zeta),
\end{equation}
where $P,r(x), \t, K(\zeta)$ are as defined in \eqref{eq:PrK}.
The left hand side and second term on the right hand side can be straightforwardly expanded using the expressions above. The integral term on the right hand side needs to be treated more carefully. Firstly, we expand using the asymptotic series,
\begin{equation}
    \frac{2}{\pi}\int_{-\infty}^0 \frac{K(\tilde{\zeta}) \tilde{\zeta}}{\zeta^2 - \tilde{\zeta}^2}d\tilde{\zeta} = \frac{2}{\pi}\int_{-\infty}^0 \frac{K_0(\tilde{\zeta}) \tilde{\zeta}}{\zeta^2 - \tilde{\zeta}^2}\de{\tilde{\zeta}}+ \mathcal{O}(\epsilon^2),
\end{equation}
where $K_0(\zeta)$ is the leading-order part of $K$ as defined in \eqref{eq:K0}. Then we apply the residue theorem to the leading-order contribution by deforming the contour to a semicircular contour in the upper-half $\zeta$-plane and noting the symmetry between the negative and positive real axes. At the first two orders the contribution from the integral cancels with the contribution from the second term on the right hand side of Bernoulli's equation \eqref{eq:bernoulliAC}. The leading-order inner equation therefore becomes,
\begin{equation}
    -\i\frac{1}{4}P^2(\zeta_*)r(b_{1*})c_{1*}\frac{\mathrm{d}^2}{\mathrm{d}z^2}\left(\nu^{-1/2}q_{in}\right) = \frac{\nu}{2c_{1*}d_{1*}r^2(b_{1*})}\left(1-\frac{1}{q_{in}^2}\right).
\end{equation}

\noindent It is natural to change variables,
\begin{equation}
    \nu = A_*^{2/7} \nu_{in} = \left(-\frac{\i P^2(\zeta_*)r^3(b_{1*})c_{1*}^2d_{1*}}{2}\right)^{2/7} \nu_{in},
\end{equation}
so then the inner equation becomes,
\begin{equation}
    - \frac{1}{\nu_{in}}\frac{\mathrm{d}^2}{\mathrm{d}\nu_{in}^2} \left(\nu_{in}^{-1/2}q_{in}\right) = \frac{1}{q_{in}} - 1,
\end{equation}
which is the same as the inner equation for the classic Saffman-Taylor problem in a channel (\cite{chapman1999}). In the outer limit (as $\tilde{\nu} \rightarrow \infty$) we thus derive the same relation as in the channel,
\begin{equation}
    q_{in} = \sum_{n=0}^\infty A_n \nu_{in}^{-7n/2},
\end{equation}
where the constants $A_n$ satisfy the recurrence relation,
\begin{subequations}
\begin{align}
    \sum_{m=1}^n \left(\frac{7m}{2}-3\right)\left(\frac{7m}{2}-2\right)A_{m-1}&\sum_{k=0}^{n-m}A_{n-m-k}A_k = \sum_{m=0}^n A_{n-m}A_m, \quad n\geq 1, \\
    A_0 &= 1.
\end{align}
\end{subequations}

\noindent Then the inner expansion gives,
\begin{equation}
    q = c_{1*}\epsilon^{-2/7}\nu^{-1/2}\sum_{n=0}^\infty A_n \nu^{-7n/2}A_*^{n},
\end{equation}
which means the $n^{\text{th}}$ term is,
\begin{equation}
 \epsilon^{2n}q_n \sim c_{1*}\epsilon^{-2/7}\nu^{-1/2}A_n \nu^{-7n/2}\left(-\frac{\i P^2(\zeta_*)r^3(b_{1*})c_{1*}^2d_{1*}}{2}\right)^n, \quad \text{as} \quad n \rightarrow \infty.
\end{equation}
Now writing the outer expansion \eqref{eq:singulantintegrated} in terms of the inner variable we find,
\begin{equation}
\begin{aligned}
    \chi &= -\int_{\zeta_*}^\zeta \frac{2\e^{\i\pi/4}\e^{-\i\tau_0/2}}{Pq_0r(x_0)^{3/2}}\de{\tilde{\zeta}} \\
    &\sim -\int_0^\nu \frac{2\e^{\i\pi/4}d_{1*}^{-1/2}\epsilon^{1/7}\tilde{\nu}^{1/4}}{P(\zeta_*)r(b_{1*})^{3/2} c_{1*}\epsilon^{-2/7}\tilde{\nu}^{-1/2}}\epsilon^{4/7}\de{\tilde{\nu}} \\
    &= \frac{8}{7} \frac{e^{-3\i\pi/4}d_{1*}^{-1/2}}{c_{1*}P(\zeta_*)r(b_{1*})^{3/2}}\epsilon \nu^{7/4},
\end{aligned}
\end{equation}
and so from the factorial over power ansatz \eqref{eq:factorialoverpower},
\begin{equation}
    \begin{aligned}
        \epsilon^{2n}q_n \sim &\Lambda_* \left(\frac{64}{49} \frac{\i}{d_{1*}c_{1*}^2 P^2(\zeta_*) r^3(b_{1*})}\right)^{-n} \nu^{-7n/2}   \left(\frac{8\e^{-3\i\pi/4}d_{1*}^{-1/2}\epsilon}{7c_{1*}P(\zeta_*)r(b_{1*})^{3/2}}\right)^{-1/14} \\
        & \times \nu^{-1/8} \e^{7\i\pi/8}2^{-1/2}d_{1*}^{-1/4}\epsilon^{-3/14} \nu^{-3/8} c_{1*}r(b_{1*})^{1/4}\Gamma(2n+\gamma),
    \end{aligned}
\end{equation}
where the $*$ subscript on the $\Lambda$ labels the corresponding singularity.
Then equating the inner and outer expansions gives,
\begin{equation}
\Lambda_* = \frac{2^{5/7}}{7^{1/14}}\e^{-13\i\pi/14}c_{1*}^{-1/14}d_{1*}^{3/14}P(\zeta_*)^{-1/14}r(b_{1*})^{-5/14} \lim_{n\rightarrow \infty} \frac{A_n}{\Gamma(2n+\gamma)}\left(\frac{32}{49}\right)^n.
\end{equation}
Using the large $n$ solution to the recurrence relation from \cite{chapman1999} we obtain a final expression for the constant,
\begin{equation}
    \Lambda_* \approx  0.46048 \times 2^{-1/7}\e^{-13\i\pi/14}\left(\frac{d_{1*}^3}{c_{1*}P(\zeta_*)r(b_{1*})^5}\right)^{1/14} .
\end{equation}

\end{document}